\documentclass[a4paper,number]{cas-sc}

\usepackage[numbers]{natbib}
\usepackage{subcaption}

\def\tsc#1{\csdef{#1}{\textsc{\lowercase{#1}}\xspace}}
\tsc{WGM}
\tsc{QE}

\begin{document}
\let\WriteBookmarks\relax
\def\floatpagepagefraction{1}
\def\textpagefraction{.001}

\usemyfooter 

\shorttitle{}    

\shortauthors{Desplenter Karel et al.}   

\title [mode = title]{CTrex: A Research‑Oriented Framework for Kernel‑ and Projection‑Level Algorithm Development in CT Reconstruction}  



%

\author[1]{Karel Desplenter}[orcid=0000-0000-0000-0000,]
\cormark[1]
\ead{karel.desplenter@ugent.be}
\credit{Software, Investigation, Writing - original draft}
\author[1]{Wannes Goethals}[orcid=0000-0002-8927-0562,]
\credit{Software, Investigation, Writing - review \& editing}

\author[1]{Thomas De Schryver}
\credit{Conceptualization, Software, Investigation, Writing - review}

\author[1]{Marjolein Heyndrickx}[orcid= 0000-0002-1967-8677,]
\credit{Software, Investigation, Writing - review}

\author[1,2]{Jan Aelterman}[orcid=0000-0002-5543-2631,]
\credit{Conceptualization, Supervision, Writing - review \& editing}

\author[1]{Matthieu N. Boone}[orcid=0000-0002-5478-4141,]
\credit{Conceptualization, Supervision, Writing - review \& editing}

\affiliation[1]{organization={Radiation Physics - UGCT, Department of Physics and Astronomy, Faculty of Science, Ghent university},
            addressline={Sint-Pietersnieuwstraat}, 
            city={Ghent},
            postcode={9000}, 
            country={Belgium}}

\affiliation[2]{organization={TELIN-IPI, Department of Telecommunications and Information Processing, Faculty of Engineering, Ghent university},
            addressline={Sint-Pietersnieuwstraat}, 
            city={Ghent},
            postcode={9000}, 
            country={Belgium}}

\cortext[1]{Corresponding author}


\begin{abstract}
Micro‑computed tomography (micro‑CT) is increasingly applied to imaging scenarios in which non‑ideal acquisition conditions such as object motion, deformation, truncated fields of view, continuous rotation, or unconventional scanning protocols violate the assumptions underlying classical reconstruction pipelines. While existing reconstruction frameworks provide efficient implementations for standard geometries and workflows, they typically expose limited control over how the forward and backward projection operators are discretized and evaluated. This limits methodological exploration when reconstruction accuracy is dominated by modelling choices at the level of ray sampling, interpolation, or operator coupling.

In this work, we introduce CTrex, a research‑oriented, GPU‑accelerated iterative reconstruction framework that exposes the full projection–correction–backprojection pipeline down to the kernel level. Rather than treating these operators as fixed black boxes, CTrex represents them as editable computational building blocks under a unified iterative control structure. This design enables reconstruction strategies that incorporate non‑ideal acquisition effects directly into the numerical operators, instead of treating them as external corrections or post‑processing steps.

CTrex combines this GPU‑level accessibility with an explicit and extensible geometry formulation based on homogeneous‑coordinate view matrices, allowing rigid and affine transformations, detector misalignments, and time‑varying acquisition geometries to be expressed consistently. This formulation can represent a broad class of scanning configurations including circular, helical, offset, and conveyor‑belt trajectories while keeping the GPU kernels agnostic to the specific trajectory definition.

A series of application studies including motion‑ and deformation‑aware reconstruction, event‑based 4D imaging, reconstruction in cylindrical coordinate domains, and extended‑field‑of‑view CT illustrate how projection‑level and kernel‑level adaptations enable reconstruction strategies that are difficult to realize in conventional frameworks. By combining explicit geometric modelling with editable operators, CTrex provides a flexible and extensible platform for advancing CT reconstruction research under realistic and unconventional imaging conditions.
\end{abstract}


\begin{highlights}
\item CTrex provides operator-level access for developing dynamic CT reconstruction methods.

\item Motion fields can be incorporated during ray traversal without full-volume warping.

\item Weighted backprojection can localise updates using prior dynamic information.

\item Angular subsampling enables motion-blur modelling in continuous-rotation acquisitions.

\item CTrex forms the operator-level basis for temporally parameterised 4D CT.

\item The framework can represent both motion and intrinsic attenuation changes.

\item Per-projection geometry accommodates non-standard and perturbed trajectories.

\item Application-specific operators retain a unified iterative control structure.

\end{highlights}


\maketitle


\section{Introduction}

Micro‑computed tomography (micro‑CT) is widely used across scientific and industrial domains, ranging from materials science and geoscience to biomedical research and non‑destructive inspection. Beyond conventional static imaging, there is growing interest in acquisition protocols for application specialized imaging geometries, that probe dynamic processes, large or partially sampled objects. In such scenarios, image quality is often limited by discrepancies between the physical acquisition process and the numerical model assumed during reconstruction.
\\\\
Non‑ideal acquisition conditions including continuous rotation during exposure, object motion or deformation, truncated fields of view, detector misalignment, and non‑standard scanning trajectories challenge the assumptions underlying many standard reconstruction pipelines. Addressing these effects frequently requires modifying how rays are traced, sampled, weighted, and accumulated, rather than adjusting only high‑level reconstruction settings such as regularization strength or strategies and iteration count. Importantly, a subset of these challenges cannot be addressed at all without direct access to the forward and inverse projection operators, as they arise from mismatches between the physical acquisition process and the numerical realization of the operators themselves. The challenges posed by these diverse use cases therefore highlight the need for reconstruction software that enables algorithmic exploration at the level of the forward and inverse operators themselves.
\\\\
A range of open‑source and commercial frameworks have been developed to support iterative CT reconstruction. Toolkits such as RTK \citep{ritReconstructionToolkitRTK2014} and ASTRA \citep{aarleFastFlexibleXray2016a, graasASTRAKernelkitGPUaccelerated2024} provide efficient GPU‑based implementations of projection and backprojection for standard geometries, while TIGRE \citep{biguriTIGREV3Efficient2025} and the Core Imaging Library \citep{jorgensenCoreImagingLibrary2021, papoutsellisCoreImagingLibrary2021} offer flexible environments for algorithm development. These frameworks are highly effective within their intended scope, prioritizing robustness, usability, and performance.
\\\\
However, this design focus typically limits direct access to the implementation and discretization of the projection operators. As a consequence, reconstruction strategies that require modifications at the operator level are difficult to express. This becomes particularly restrictive in applications where the reconstruction model must be adapted directly to the acquisition process, such as motion compensation, deformation‑aware reconstruction, non‑Cartesian reconstruction domains, or extended fields of view. In practice, these problems are often approximated through indirect workarounds, for example by repeatedly resampling intermediate volumes, introducing auxiliary reconstruction stages. Such approaches increase computational cost, complicate implementation, and obscure the physical interpretation of the reconstruction model.
\\\\
In this work, we address this limitation by presenting CTrex, an extensible reconstruction framework designed explicitly for reconstruction algorithm research. Rather than supporting a fixed set of geometries or workflows, CTrex exposes projection, correction, and backprojection operators as editable GPU kernels within a unified iterative control structure. This projection‑level accessibility enables reconstruction strategies to be modified directly at the level where physical modelling and numerical discretization interact, enabling controlled investigation of operator‑level design choices.
\\\\
A key enabling component of the framework is its explicit geometry formulation, based on homogeneous‑coordinate view matrices that describe the relative configuration of source, detector, and volume. These parameters are defined at the framework level and provided to the GPU kernels on a per‑projection basis, allowing the kernels themselves to remain general ray‑integration operators. This separation makes it possible to represent rigid and affine transformations, detector misalignment, and time‑varying acquisition geometries, including circular, helical, offset, and conveyor‑belt trajectories to be expressed within a consistent operator formulation, without redesigning the kernel implementations.
\\\\
In addition to methodological flexibility, explicit operator formulations allow performance‑critical sections of the reconstruction to be optimized without sacrificing model fidelity. Because numerical integration, sampling, and interpolation are implemented explicitly in GPU kernels, computational optimizations can be applied in a targeted manner, informed by both the physical model and the hardware architecture. This positions CTrex as a suitable testbed for exploring reconstruction strategies that balance physical realism, computational cost, and application‑specific requirements.
\\
The remainder of this paper is structured as follows. Section~\ref{sec:basic_concept} introduces the overall architecture of CTrex, including the reconstruction control, geometry handling, and GPU kernel execution. Section~\ref{sec:application_study} presents a series of application studies demonstrating how operator‑level access enables reconstruction tasks that are difficult to realize in conventional frameworks. Subsection~\ref{sec:kernel_level_adap} provides an introductory description of the baseline projection, correction, and backprojection operators, followed by the kernel‑level adaptations that supported these application studies. Section~\ref{sec:discussion} discusses methodological implications and limitations, and Section~\ref{sec:conclusion} concludes the paper and outlines directions for future work.

\section{Basic concept and framework design}\label{sec:basic_concept}

To enable reproducible algorithmic research and systematic exploration of non-standard reconstruction strategies, we developed CTrex as an open, extensible reconstruction environment that exposes the internal stages of the iterative reconstruction pipeline. While several methodological contributions based on CTrex have been reported in earlier work, this paper provides the first consolidated description of the framework itself, with a focus on its architectural principles and implementation choices that support kernel-level algorithm development.
\\\\
Rather than presenting a monolithic reconstruction application, CTrex is structured to separate reconstruction configuration, data management, and GPU kernel execution. This separation allows researchers to modify and extend individual components of the reconstruction pipeline such as geometry handling, projection models, or update strategies without disrupting the overall workflow.
\\\\
This section therefore provides an overview of the core architectural concepts of CTrex and the software abstractions that define its reconstruction workflow. We first describe the overall system architecture and its main components, followed by the geometric model and trajectory representation used to parameterize forward and backward operators. The section concludes with a description of the \texttt{Rex} class, which controls iteration logic and operator scheduling during reconstruction.

\subsection{Architecture}

\begin{figure}
    \centering
    \includegraphics[width=0.35\textwidth]{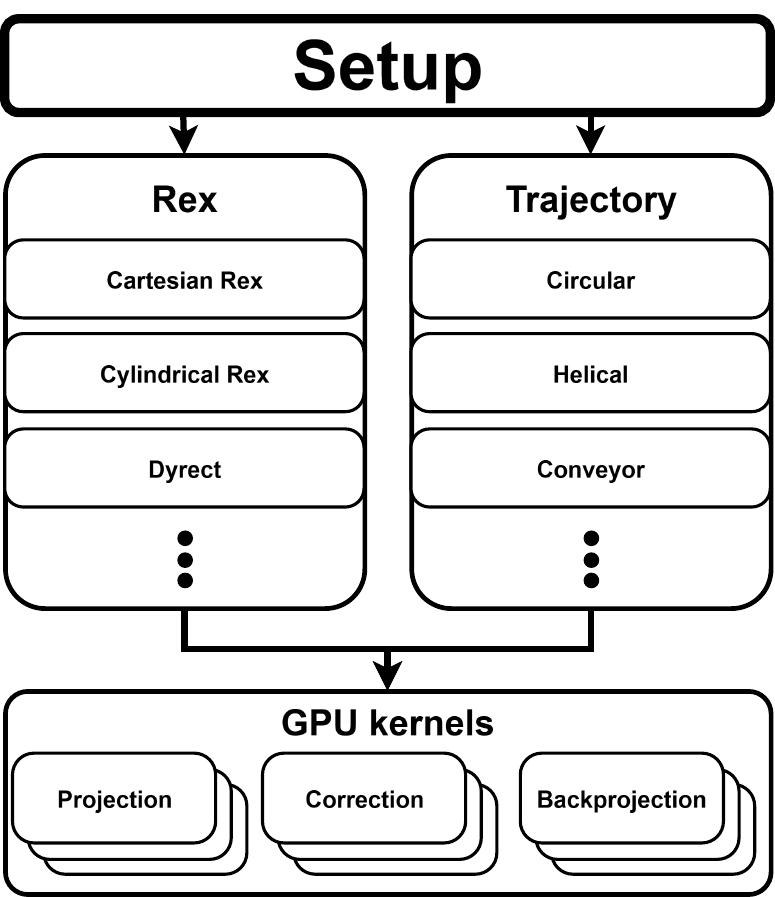}
    \caption{Overview of the CTrex reconstruction architecture. The user-defined configuration is passed to the Setup component, which initializes the appropriate Rex and Trajectory classes. During reconstruction, GPU kernels receive per-view geometry information from the Trajectory object, while execution order and iteration logic are controlled by the Rex class.}
    \label{fig:arch3}
\end{figure}

The reconstruction framework is organized around a small set of clearly separated components, each with a well‑defined responsibility. An overview of the interaction between these components is given in Fig.~\ref{fig:arch3}. The design allows the reconstruction workflow to be adapted to different acquisition geometries and reconstruction algorithms through the selection of appropriate \texttt{Rex} and \texttt{Trajectory} implementations.
\\\\
The reconstruction is initialized through a dedicated \texttt{Setup} class, which is responsible for configuring the computational environment and providing a consistent system description. This includes loading input data, initializing acquisition and reconstruction geometry and enabling or disabling optional processing steps. Operations such as preprocessing or correction of projection data are not hard‑coded into the setup phase, but are configured/modified as optional stages that can be applied or bypassed depending on the reconstruction task.
\\\\
Once the configuration is complete, the reconstruction process is controlled by an instance of a \texttt{Rex} class. The \texttt{Rex} abstraction represents a specific reconstruction algorithm and defines how projection, correction, and backprojection operations are combined and iterated. Based on the system description provided by \texttt{Setup}, the \texttt{Rex} instance prepares all data structures required for execution, including geometry parameters, projection data, and volume buffers, and manages their transfer to GPU memory. It further defines the execution order of the computational kernels, controls iteration counts, implements ordered‑subset or similar iteration schemes, and ensures synchronization between successive kernel launches.
\\\\
Geometrical information required by the reconstruction kernels is provided by a separate \texttt{Trajectory} class. This component describes the acquisition path and the spatial relationship between source, detector, and reconstruction volume. The \texttt{Trajectory} is independent of the reconstruction algorithm itself and supplies view‑dependent geometry parameters to the GPU kernels during execution.
\\\\
The interaction with the GPU is abstracted through a Python‑to‑GPU interface that provides device memory management, runtime kernel compilation, and kernel execution. While the \texttt{Rex} class determines which kernels are executed and how they are parameterized, the low‑level compilation and launch of these kernels is handled by this interface. In the present implementation, this functionality is realized using PyCUDA, which enables runtime generation and specialization of GPU kernels directly from Python, as well as explicit control over data transfer between host and device. Although this implementation targets CUDA‑capable hardware, the architectural separation between reconstruction control and kernel execution is not tied to a specific GPU vendor, allowing alternative backend interfaces to be substituted without changes to the high‑level reconstruction logic. This separation allows reconstruction algorithms to be modified and extended at a high level, while retaining the performance characteristics of native GPU implementations.

\subsection{Geometric Model and Trajectory Representation}

All numerical reconstruction operators in CTrex act on a per‑projection geometric description that defines how detector measurements are mapped into the reconstruction volume. Rather than encoding scan motion or acquisition strategy directly inside the GPU kernels, CTrex separates the definition of geometrical relationships from their numerical evaluation. This separation allows the same projection and backprojection operators to be reused across a wide range of acquisition configurations by updating only the per‑projection geometry parameters.

\subsubsection{Single‑projection geometry}

For each projection image, the reconstruction operators operate on a geometric description expressed in the volume‑attached reference frame. This description specifies the relative configuration of source, detector, and reconstruction volume and is provided to the GPU kernels in the form of homogeneous view matrices and a corresponding $3 \times 4$ projection matrix.
\\\\
Two coordinate frames are used. The laboratory reference frame \textbf{L} is a fixed external coordinate system in which the physical scanner configuration and any object motion are defined. Reconstruction is performed in the volume‑attached frame \textbf{V}, which serves as the computational reference frame for forward and backprojection. Prior to kernel execution, all geometry parameters are transformed from \textbf{L} to \textbf{V}, and only this relative geometry is passed to the GPU.
\\\\
Within this formulation, the X‑ray source is modeled as a point source described by its three‑dimensional position, while the detector is described by a full six‑degree‑of‑freedom pose. The reconstruction volume itself does not appear explicitly in the GPU interface; instead, its pose is absorbed into the transformation between frames \textbf{L} and \textbf{V}. As a result, each projection is fully characterized by nine degrees of freedom (DOF): three for the source position (3 DOF) and six for the detector pose (6 DOF), both expressed in frame \textbf{V}.
\\\\
Encoding geometry in this relative form ensures that the numerical projection and backprojection operators remain independent of scanner layout, mechanical design, or acquisition protocol.

\subsubsection{Trajectory representation}

A trajectory is defined as the ordered sequence of per‑projection geometries describing how the relative configuration of source, detector, and volume evolves across projection images. In CTrex, trajectories are not implemented through specialized kernel logic. Instead, they are represented implicitly by updating the view matrices supplied to the kernels for each projection.
\\\\
This design renders the reconstruction operators entirely agnostic to the acquisition trajectory. Circular, helical, offset, and non‑standard scan paths differ only in the sequence of geometric parameters provided to the kernels. Changes in scanner motion, acquisition strategy, or projection ordering therefore require no modification of the numerical operators themselves.
\\\\
Common trajectories arise as specific parameterizations of this general formulation. For example, a simple helical trajectory can be typically described by a rotation
$\mathbf{R}(t) = \mathbf{R}_z(\alpha(t))$
about a fixed axis combined with a translation
$\mathbf{T}(t) = (0, 0, p\,\alpha(t)),$
where $p$ denotes the helical pitch expressed as translation per unit rotation angle ($\alpha(t)$ the rotation angle at $t$), a special case $p = 0$ corresponds to the circular scan. More generally, both the rotation
$\mathbf{R}(t) = \mathbf{R}_T(\alpha(t))$
 and translation 
 $\mathbf{T}(t) = (x(t), y(t), z(t))$
are not restricted to a fixed axis or linear relation, but may follow arbitrary time‑dependent transformations, allowing the motion axis and trajectory to vary throughout the acquisition. This enables  unconventional acquisition schemes, such as conveyor‑belt geometries in which the object translates and rotates while the source–detector pair remains fixed, to be expressed as reparameterized sequences of affine per‑projection transformations \citep{deschryverInlineNondestructiveEvaluation2017}.

\paragraph{Affine volume pose}

The mapping between the laboratory frame \textbf{L} and the volume frame \textbf{V} is expressed as a general affine transformation. This representation encompasses rigid motion as well as scaling and shear, allowing calibration errors, geometric distortions, or controlled deformations to be incorporated directly into the reconstruction model.
\\\\
In practice, the affine pose is represented by a $4 \times 4$ homogeneous transformation matrix acting on ray coordinates. For numerical stability and interpretability, this transformation is internally decomposed into a rigid component followed by an affine component. During ray construction in the GPU kernels, the rigid transformation establishes the geometric relationship between source, detector, and volume, while the affine component modifies the metric properties of the ray without altering its origin.
\\\\
By expressing both motion and deformation through affine per‑projection transformations, CTrex provides a unified geometric model in which reconstruction kernels operate on the most general representation possible, independent of scanner realization or acquisition strategy.

\subsection{Rex class}

The \texttt{Rex} class implements and controls the iterative reconstruction process. It defines how projection data is selected, how numerical operators are sequenced, and how volume updates are applied across iterations. Rather than encoding individual reconstruction algorithms as fixed routines, \texttt{Rex} provides a general iteration framework in which different update strategies are realized through configuration and indexing.

\subsubsection{Iterative reconstruction loop}\label{sec:iterative_schemes}

The reconstruction methods implemented in CTrex follow a general ordered‑subset SIRT formulation, from which classical SIRT and SART arise as specific cases. These variants share a common update structure and differ only in how projection data are grouped and scheduled during each iteration update step.
\\\\
For a static object, the attenuation volume ($\mu$) update at iteration index $k$ can be written as
\begin{equation}\label{eq:sirt}
\mu_{k+1} =
\underbrace{\mu_{k} + \alpha \, C A^T 
\underbrace{R\left( p - \underbrace{A\mu_k}_{1}\right)}_{2}}_{3}
\
with \ k \in [1,\ \text{iteration end}]
\end{equation}
where $A$ denotes the forward projection operator, $A^T$ the corresponding backprojection, $p$ the measured projection data, $\alpha$ a relaxation parameter, and $C$ a diagonal voxel‑space normalization operator.
\\\\
In this formulation, $A\mu_k$ [marked as \textbf{1} in Eq.~\eqref{eq:sirt}] represents the simulated projections generated from the current volume estimate. The difference $(p - A\mu_k)$ is computed in detector space and optionally weighted by $R(\cdot)$ [marked as \textbf{2}], which incorporates geometric weighting, statistical weighting, or ordered‑subset selection. 
The weighted residual is then backprojected by $A^T$, normalized by $C$, and scaled by the relaxation parameter $\alpha$ to form the volume update [marked as \textbf{3}], yielding the next estimate $\mu_{k+1}$. 
\\\\
The role of the \texttt{Rex} class is to orchestrate this update loop: selecting the projection data contributing to $A$ and $A^T$ at each iteration, applying the appropriate weighting operators $R$ and $C$, and advancing the iteration counter. In this section, the focus is on how these operators are scheduled and combined during reconstruction; their operator‑level implementation is described later in the technical description of kernel‑level operators Sec.~\ref{sec:kernels}.

\subsubsection{Projection indexing and ordered‑subset control}\label{sec:indexing_os}

Ordered‑subset methods are realized in CTrex by selecting, at each iteration, which projections contribute to the evaluation of Eq.~\eqref{eq:sirt}. To this end, \texttt{Rex} maintains an explicit indexing scheme that defines subsets of projection indices used during reconstruction.
\\\\
Let $\mathcal{P}_k \subset \{1,\dots,N_\text{proj}\}$ denote the set of projection indices selected at iteration $k$. The forward projection, correction, and backprojection operations of equation~\eqref{eq:sirt} are then evaluated using only 
the views in $\mathcal{P}_k$, i.e. $A(\mu, \mathcal{P}_k)$, $R(\cdot, \mathcal{P}_k)$, and $A^{T}(\cdot, \mathcal{P}_k)$. The choice of $\mathcal{P}_k$ is configurable at runtime and may be deterministic, cyclic, or randomized, optionally incorporating geometry‑dependent ordering or weighting.
 Different iterative schemes are obtained through the choice of the sequence $\{\mathcal{P}_k\}$: selecting a single projection per update yields SART‑like behaviour; grouping all projections into a single subset recovers classical SIRT; intermediate groupings correspond to OS‑SIRT methods.
\\\\
Because projection selection is handled at the indexing level, changes in subset size or ordering do not require modifications to the kernel implementations or recompilation of GPU code. By separating projection selection from the numerical operators, the \texttt{Rex} class enables systematic investigation of update ordering, subset size, and weighting strategies within a unified reconstruction framework.

\section{Application study}\label{sec:application_study}

CTrex has been serving as the foundation for developing multiple application‑specific reconstruction pipelines across a range of micro‑CT use cases. This section presents selected case studies that illustrate how the framework accommodates varying acquisition conditions, motion models, and reconstruction constraints by adapting the projection and backprojection operators at the kernel level.
\\\\
These examples demonstrate how operator‑level flexibility enables practical reconstruction strategies for non‑ideal imaging conditions. Each case highlights a specific limitation of conventional approaches and shows how it can be addressed through targeted modifications of the numerical operators. The corresponding implementation details are discussed in Section~\ref{sec:kernel_level_adap}

\subsection{Motion compensation warping using Digital volume correlation (DVC) displacement fields}\label{sec:DVC}

\begin{figure}
    \centering
    \includegraphics[width=0.85\textwidth]{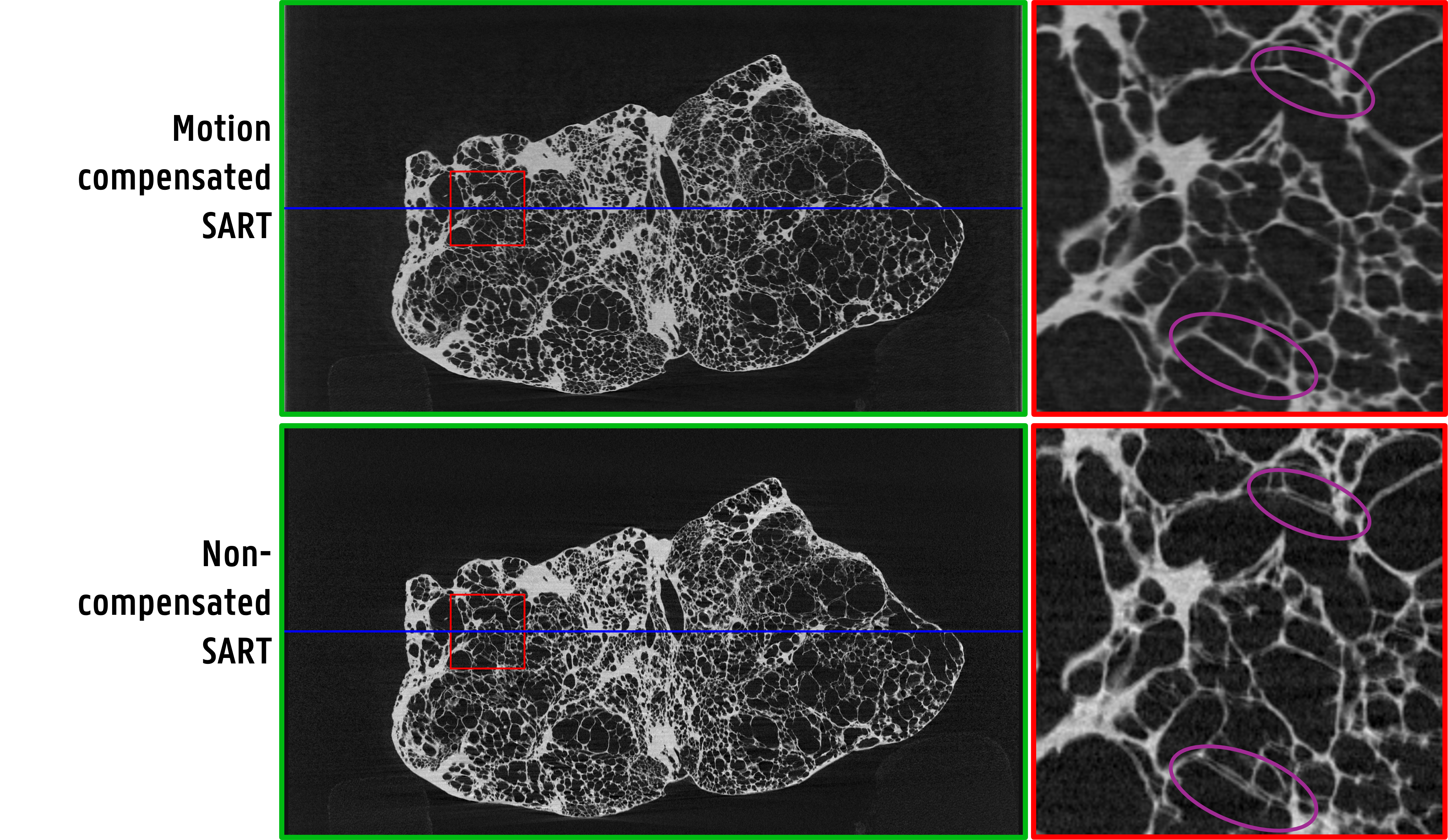}
    \caption{CTrex‑enabled motion‑compensated reconstruction applied to a micro‑CT acquisition of lungs. Comparison between reconstructions with motion compensation (\textbf{top row}) and without motion compensation (\textbf{bottom row}). The right column shows a zoom‑in of the reconstructed slice where motion‑induced artefacts in the uncompensated reconstruction lead to apparent duplication of thin structures (indicated by circles), which is suppressed when deformation is incorporated directly into the projection and backprojection operators.} 
    \label{fig:motion-lungs}
\end{figure}

Our first application study concerns motion‑compensated reconstruction using displacement fields derived from digital volume correlation (DVC), with the aim of mitigating artefacts caused by sample motion during acquisition.  Conventional reconstruction algorithms assume that the sample remains static throughout data acquisition. When this assumption is violated, inconsistencies arise between projections, leading to streaking, blurring, and reduced spatial resolution in the reconstructed volume. In micro‑CT, such inconsistencies may result from sample deformation, mechanical drift, or dynamic in‑situ processes. In four‑dimensional CT, temporal sample evolution is intrinsic to the problem formulation and must therefore be explicitly accounted for during reconstruction.
\\\\
Addressing these effects requires a reconstruction approach in which motion is incorporated consistently into the imaging model. A straightforward (naive) approach is to deform the reconstruction volume separately for each projection and then apply a conventional static projection operator, effectively treating motion as a sequence of volume transformations. Although conceptually simple, this approach does not integrate the motion model directly into the operator evaluation and requires repeated interpolation of the full volume.
\\\\
A more consistent formulation is to incorporate motion directly into the projection and backprojection operators, allowing the imaging model to remain consistent with the acquisition process. CTrex enables this by providing direct, programmable access to the forward‑ and backprojection kernels, so that sample‑space transformations such as deformation fields can be integrated explicitly during ray traversal. Within this framework, the displacement field used for motion compensation can be obtained from a Digital Volume Correlation (DVC) model, which estimates a three‑dimensional displacement field by correlating a reference volume with a deformed configuration \citep{quintanaDigitalVolumeCorrelation}. This allows both rigid motion and non‑rigid deformation to be incorporated into the reconstruction process \citep{DVC, goethalsDVCDynamicCT2019, manigrassoMultiframeDVCTemporal2022, deschryver4DVizualisationInsitu2017}. The same formulation can be applied both to  time‑resolved acquisitions and to nominally static scans affected by intra‑scan motion, thereby improving reconstruction quality in both 3D and 4D settings.
\\\\
An example of improved reconstruction quality is shown in Fig.~\ref{fig:motion-lungs}, illustrating a nominally static micro‑CT scan of lung tissue \citep{VERLEDEN2021167} affected by intra‑scan motion. The sample, consisting of inflated frozen lungs maintained on dry ice, exhibits gradual motion during acquisition. A motion model was estimated by comparing projections at $0^\circ$ and $360^\circ$ and linearly interpolating the displacement over the acquisition. In the uncompensated reconstruction, this motion results in typical duplication artefacts in thin structures, while these artefacts are suppressed when the deformation model is incorporated directly into the reconstruction.
\\\\
The deformation is represented by dense displacement fields $\mathbf{u}_p(\mathbf{x})$, defined for each projection image $P_p$, describing the sample motion at the time of acquisition.
These displacement fields, typically obtained using DVC, describe the motion vector field between a reference state and a deformed configuration. They are typically defined at discrete time points and interpolated to obtain projection‑specific displacement fields $\mathbf{u}_p(\mathbf{x})$ corresponding to each acquisition angle. Within CTrex, these motion vector fields can be integrated directly into the projection operator by extending the ray‑sampling logic in the projector kernel (Sec.~\ref{sec:P_kern_DVC}).

\subsubsection{Reconstruction strategy}\label{sec:DVC_recon}

\begin{figure}
    \centering
    \includegraphics[width=\textwidth]{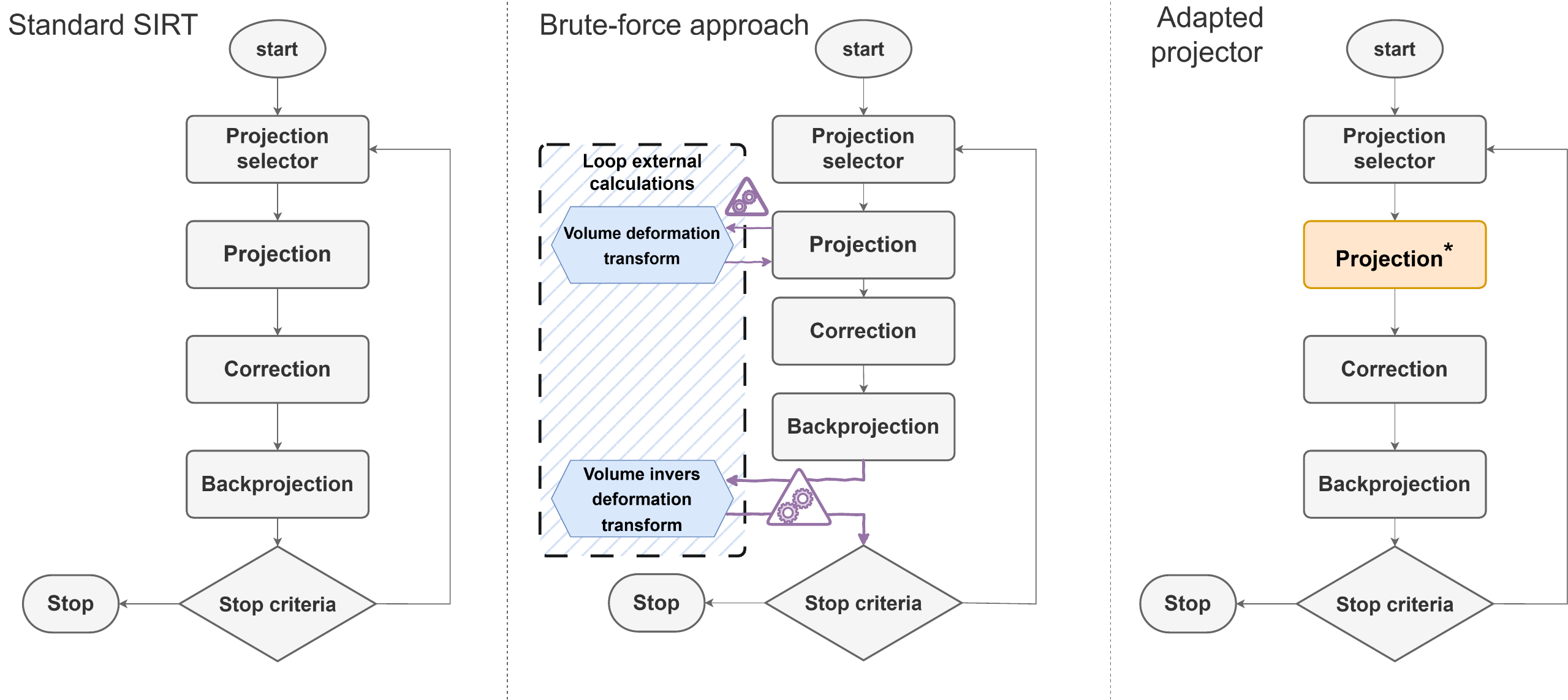}
    \caption{Comparison of SIRT iteration loop structures. Shown are the standard SIRT formulation (\textbf{left}), a brute‑force (\textbf{middle}) DVC approach incorporating explicit volume deformation and inverse deformation between projection and backprojection, and the CTrex implementation (\textbf{right}) using an adapted projection operator ($\mathrm{projection}^*$) in which deformation is integrated directly into the kernel via an additional texture lookup \citep{DVC}.}
    \label{fig:DVC-sirt-loop}
    
\end{figure}

To illustrate the role of operator‑level integration, Fig.~\ref{fig:DVC-sirt-loop} (\textbf{middle}) compares a naive motion-compensation strategy  with the kernel‑integrated approach implemented in CTrex (Fig.~\ref{fig:DVC-sirt-loop} \textbf{right}). In a naive (brute‑force) implementation, the deformation field is applied to the entire intermediate volume for each projection, followed by projection, backprojection, and inverse warping. While conceptually straightforward, this requires repeated resampling of the full volume for every projection and iteration, making it computationally impractical for iterative schemes such as OS‑SIRT.
\\\\
In contrast, CTrex integrates the deformation directly into the ray‑sampling step of the projector and, where needed, the backprojector kernels. During ray traversal, sample coordinates are mapped through the projection‑specific displacement field, such that each ray samples the volume at its deformed location while remaining expressed in a common reference coordinate system. The motion vector fields are thus evaluated per ray only at the sample locations required during operator execution, rather than applied as explicit transformations to the entire volume. This operator‑level formulation limits interpolation to a single operation per ray and avoids repeated resampling of the full reconstruction volume across projections and iterations.
\\\\
As a result, the kernel‑integrated formulation can provide a more efficient and more stable representation of motion within iterative reconstruction. By incorporating deformation directly into the operator evaluation, it reduces interpolation artefacts associated with repeated full‑volume resampling and enables motion compensation without transforming the entire reconstruction volume at each iteration (Fig.~\ref{fig:DVC-sirt-loop}).
\\\\
This application illustrates how motion compensation can be incorporated into the reconstruction loop through a limited modification of the projection operator, without altering the overall iteration structure controlled by the \texttt{Rex} class. As a result, motion‑aware reconstruction can be performed with a computational cost comparable to conventional iterative reconstructions on static samples, while significantly improving robustness to deformation.

\subsection{Weighted backprojection}\label{sec:weight}

In fast dynamic micro‑CT acquisitions, reduced exposure time and limited angular sampling typically lead to elevated noise levels. Standard iterative reconstruction schemes tend to propagate noise and motion‑induced inconsistencies throughout the entire volume, including regions that are expected to remain static during the experiment.
\\\\
Weighted backprojection addresses this limitation by incorporating prior knowledge about the expected spatiotemporal behavior of the sample \citep{heyndrickxImprovingImageQuality2020a, WBP}. This prior information commonly consists of (i) a reference volume approximating the sample structure before or after the dynamic process and (ii) a spatial estimate of which regions are likely to undergo change during acquisition \citep{heyndrickxImprovingReconstructionDynamic2016}. Rather than enforcing this information globally, weighted backprojection introduces it locally and gradually during reconstruction.
\\\\
In this formulation, each voxel $j$ is assigned a continuous weight $w_j$ that reflects its likelihood of change. Voxels in dynamic regions receive higher weights and therefore absorb a larger fraction of the projection‑domain correction, while static regions receive lower weights and remain closer to their reference values. This selective update mechanism reduces noise propagation into static regions and accelerates convergence in areas where temporal variation is expected, without requiring the prior information to be exact.
\\\\
In CTrex, weighted backprojection is implemented directly within the backprojection kernel (Sec.~\ref{sec:B_kern_weight}), allowing voxel‑wise weights to modulate the accumulation of correction values during backprojection.
Because projection‑domain corrections are concentrated in dynamic regions, these voxels may experience increased sensitivity to noise. This effect is mitigated through the use of non‑binary weighting and by controlling the influence of each backprojection update at the voxel level.
As a result, dynamic regions absorb a larger fraction of the update, while static regions remain more weakly coupled to the projection residuals.
\\\\
Embedding the weighting logic directly into the backprojection operator avoids explicit full‑volume correction steps between iterations. In conventional implementations without operator‑level access, weighted updates would typically require constructing masked or weighted correction volumes and applying additional full‑volume passes at each iteration, increasing memory traffic and computational overhead in proportion to the reconstruction volume size. By incorporating the weighting directly into the backprojection kernel, CTrex enables spatially localized use of prior information while preserving the global iteration structure managed by the \texttt{Rex} class and maintaining consistency between the forward and backprojection operators.

\begin{figure}
    \begin{subfigure}[c]{0.329\textwidth}
    \includegraphics[width=0.9\textwidth]{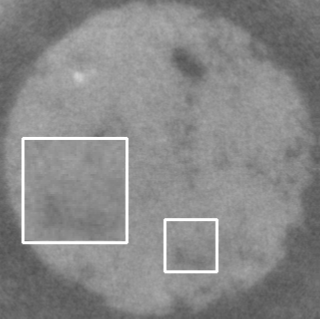}
    \caption{SART reconstruction.}
    \label{subfig:exp_WBP_normal}
    \end{subfigure}
    \begin{subfigure}[c]{0.329\textwidth}
        \includegraphics[width=0.9\textwidth]{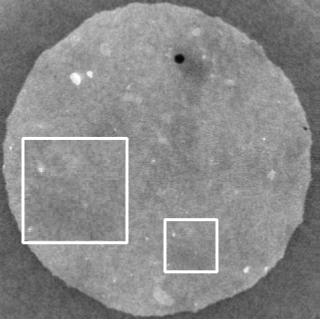}
        \caption{Reconstruction from an initial volume.}
        \label{subfig:exp_WBP_init}
    \end{subfigure}
    \begin{subfigure}[c]{0.329\textwidth}
        \includegraphics[width=0.9\textwidth]{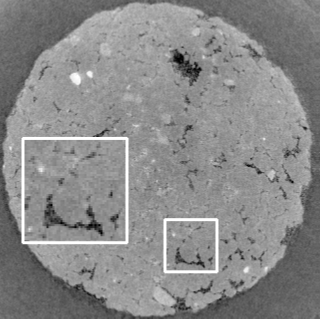}
        \caption{Weighted backprojection reconstruction.}
        \label{subfig:exp_WBP}
    \end{subfigure}
     \caption{4D‑$\mu$CT of fluid flow through Bentheimer sandstone \citep{WBP}. 
    Compared to conventional SART reconstruction (\textbf{a}), voxel‑wise weighted backprojection implemented directly inside the CTrex backprojection kernel (\textbf{c}) suppresses noise propagation into regions expected to remain static by modulating correction accumulation at the operator level. This leads to improved reconstruction quality compared to reconstruction initialised from a reference volume alone (\textbf{b}).}
    \label{fig:exp_WBP}
\end{figure}

\subsection{4D event reconstruction}

While motion‑compensated reconstruction can restore geometric consistency in the presence of sample displacement or deformation, many dynamic micro‑CT experiments involve intrinsic changes in attenuation, where local attenuation evolves irreversibly over time. Examples include phase transitions, fracture initiation, or fluid invasion, in which attenuation changes cannot be described as spatial transformations of a static structure. In such cases, representing dynamics purely through motion models is insufficient to capture the underlying physical processes and motivates reconstruction strategies in which temporal evolution is explicitly incorporated into the inverse problem. Addressing these effects requires extending the forward and inverse operators to account for time‑dependent attenuation, which is enabled in CTrex through direct access to the projection and backprojection kernels.
\\\\
Conventional 4D CT typically represents dynamics by reconstructing a sequence of quasi‑static 3D volumes from successive subsets of projections, each acquired over a finite time interval. In this frame‑based formulation, temporal resolution is fundamentally limited by the acquisition time required to collect a sufficient number of projections per frame. Moreover, correlations between projections acquired at different time points are not explicitly exploited, despite the inherent redundancy present in the data.
\\\\
Event‑based 4D reconstruction addresses these limitations by parameterising temporal behaviour at the voxel level, rather than reconstructing a sequence of independent frames. Instead of grouping projections into discrete time intervals, the forward model evaluates projection values based on temporally parameterised attenuation, allowing residuals from individual projections to directly inform the estimation of temporal evolution. Within CTrex, such approaches are implemented by jointly estimating a small set of temporal parameters per voxel from the full projection dataset, as demonstrated by the DYRECT method \citep{dyrect}. In this formulation, temporal evolution is represented using three parameter fields: an initial attenuation volume, a final attenuation volume, and a transition‑time map encoding when local changes occur. These parameters are estimated simultaneously within a single iterative reconstruction procedure, yielding a compact representation that avoids reconstructing intermediate time frames and reduces dependence on frame‑based temporal discretisation.
\\\\
Implementing event‑based reconstruction schemes of this kind requires update rules that differ fundamentally from classical SIRT or SART formulations. Projection residuals must be evaluated with respect to temporally parameterised voxel models, and backprojection updates must act on multiple parameters per voxel rather than on a single attenuation value. As a result, both the forward and inverse operators must be extended to incorporate temporal behaviour directly during projection and backprojection. The kernel‑level adaptability of CTrex provides the required flexibility to implement such temporally parametric reconstruction strategies within a unified iterative control structure.

\begin{figure}
    \centering
    \includegraphics[width=0.5\textwidth]{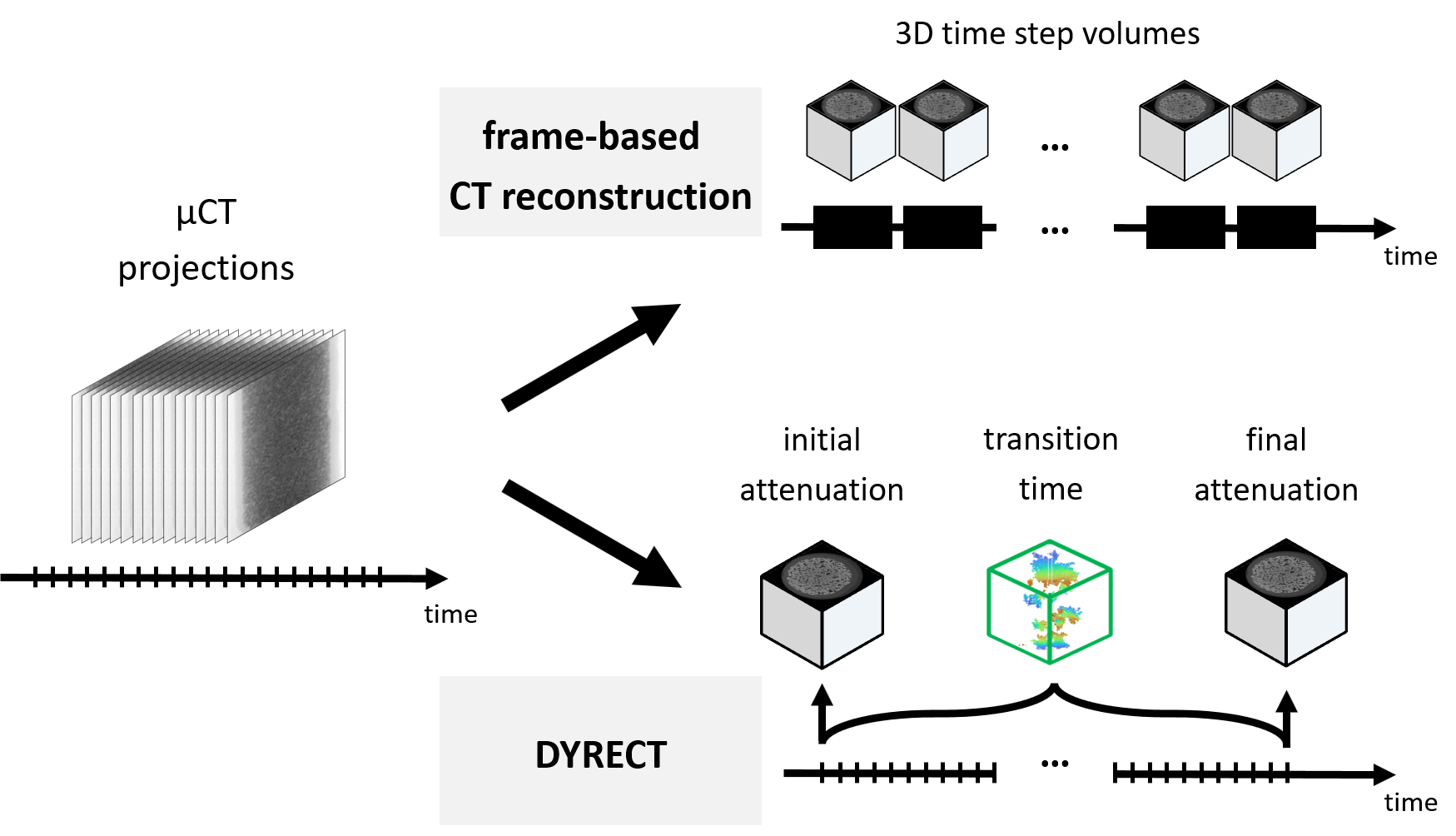}
    \caption{Schematic comparison between the event-based reconstruction (DYRECT) and conventional frame-based reconstruction techniques.
DYRECT describes events with fine temporal resolution at the projection level,
using a single volume of local transition times. This mitigates temporal blurring associated with the longer frame sequence of coarse-resolution time step
volumes. \citep{dyrect}}
    \label{fig:DYRECT}
\end{figure}

\subsection{Motion blur simulation}\label{sec:motion}

CTrex enables explicit modelling of angular motion blur at the level of the forward projection operator, allowing continuous‑rotation (fly‑scan) CT acquisitions to be reconstructed in a manner consistent with the physical image‑formation process. In such acquisitions, the rotation stage moves continuously during exposure, causing each projection to represent an angular integration rather than an instantaneous measurement \citep{cantModelingBlurringEffects2015}. Most reconstruction algorithms, however, assume a static object pose, introducing a mismatch with the physical image formation process.
\\\\
As a result, part of the correction computed during iterative reconstruction compensates for modeling error in the projector rather than true discrepancies between the current volume estimate and the measured data. Even for an otherwise accurate volume representation, residual artefacts may persist due to the unmodeled temporal averaging inherent to fly scanning.
\\\\
Within CTrex, this effect can be addressed by explicitly incorporating angular motion blur into the forward projection model. In frameworks with fixed projection models, such effects are typically approximated through ad‑hoc deblurring or external correction steps, as incorporating angular integration directly into the forward model would require modifying the projector implementation itself. Rather than treating motion blur as an external artefact, in CTrex it can be modeled as part of the image formation process by extending the pixel‑driven projector with an additional integration dimension. Each detector pixel value is approximated by accumulating contributions from multiple sub‑projections corresponding to discrete angular offsets along the continuous rotation trajectory. In the kernel, this is implemented as an additional loop over sub‑projection angles in the projection kernel.
\\\
In the corresponding backprojection step (Sec.~\ref{sec:B_kern_motion}), the correction should reflect the same angular integration used in the forward model. Rather than treating each projection as a single measurement, the residual can be distributed over the contributing sub‑projection angles, ensuring matched temporal discretisation between forward and backward operators. These modifications improve the consistency of the imaging model without altering the global iteration structure managed by the \texttt{Rex} class.

\begin{figure}
    \centering
    \includegraphics[width=0.95\textwidth]{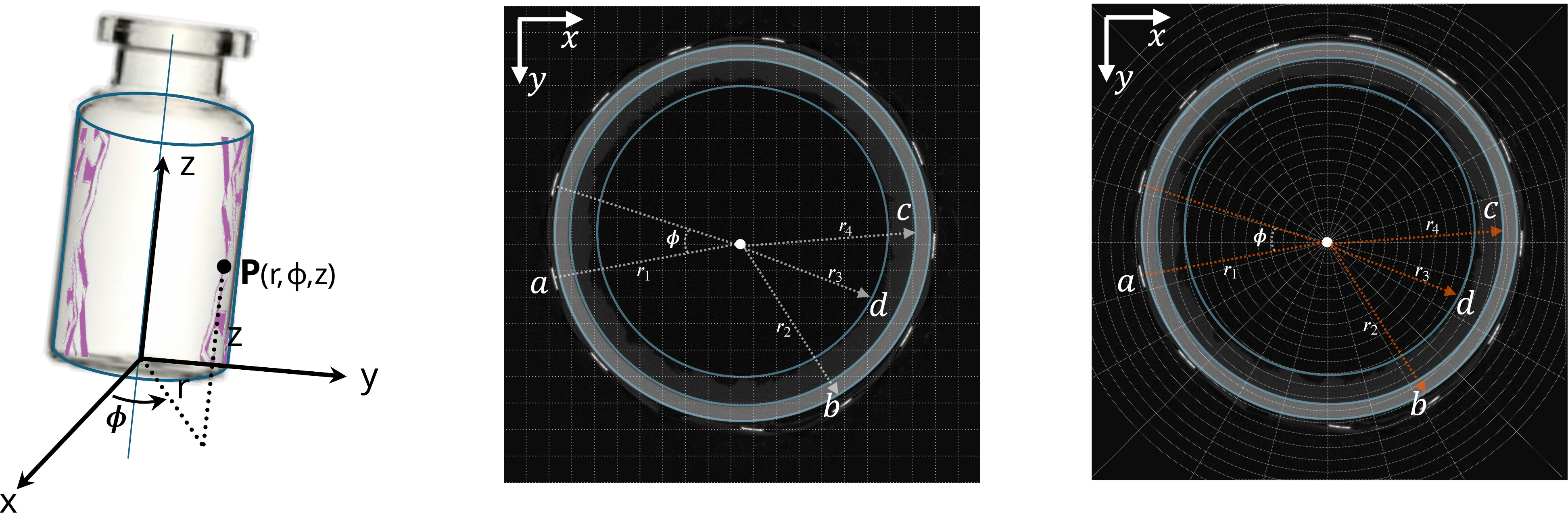}
    \caption{Example of reconstruction in aligned cylindrical coordinates of a spin‑frozen solution in a Schott vial (\textbf{left}), together with a schematic overlay of Cartesian (\textbf{middle}) and cylindrical (\textbf{right}) voxel grids on a reconstructed slice. In the Cartesian representation, concentric sample features are misaligned with the voxel grid, necessitating interpolation for cylindrical analysis, whereas they are naturally aligned in the cylindrical grid centred on the object's symmetry axis. Reconstruction directly in cylindrical coordinates enables concentric structures to be represented and segmented without Cartesian‑to‑cylindrical re‑interpolation \citep{goethalsInsituXrayImaging2020}.}
    \label{fig:exp_cylinder}
\end{figure}

\subsection{Aligned cylindrical coordinates}\label{sec:cylinder}
An example of reconstruction directly in cylindrical coordinates is shown in Fig.~\ref{fig:exp_cylinder}, where concentric structures are represented and segmented without requiring Cartesian‑to‑cylindrical re‑interpolation \citep{goethalsInsituXrayImaging2020}. This allows geometric features that naturally follow cylindrical symmetry to be preserved throughout both reconstruction and downstream analysis.
\\\\
CTrex enables reconstruction directly in non‑Cartesian volume domains when the sample geometry is known a priori, which is advantageous for imaging problems where structural features follow a well‑defined coordinate system \citep{goethalsStrategiesConeBeam2019, cylindric, goethalsInsituXrayImaging2020}. This is particularly relevant in high‑throughput inspection workflows, where downstream analysis operates in non‑Cartesian coordinates and conventional Cartesian reconstructions require coordinate transformations that introduce interpolation error and additional processing overhead.
\\\\
One representative case concerns the inspection of cylindrical or tube‑like components. In such scenarios, tasks such as segmentation, dimensional measurement, or defect detection are naturally expressed in cylindrical coordinates. Reconstructing directly in a cylindrical volume domain preserves geometric alignment between the reconstruction and the analysis space, avoiding inconsistencies introduced by post‑hoc coordinate transformations.
\\\\
In CTrex, this is realized by implementing projector and backprojector kernels that operate on an aligned cylindrical volume representation (Secs.~\ref{sec:P_kern_cyl} and \ref{sec:B_kern_cyl}). While the underlying ray integration remains based on linear interpolation along discretised ray samples, the coordinate mapping and voxel indexing are defined in cylindrical coordinates rather than Cartesian space. This requires direct control over the kernel implementation, as the mapping between physical space and voxel indices differs fundamentally from standard Cartesian grids, while the surrounding reconstruction algorithm remains unchanged.

\begin{figure}
    \centering
    \includegraphics[width=0.75\textwidth]{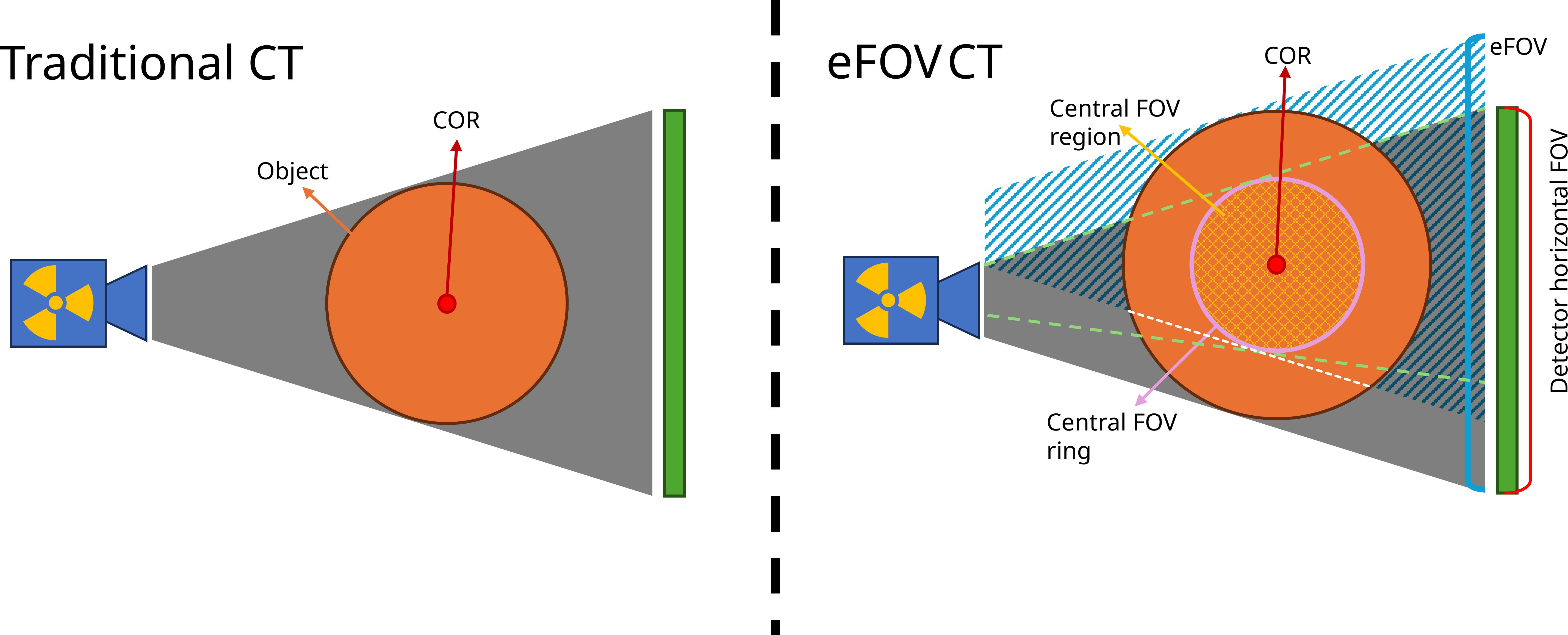}
    \caption{Unconventional acquisition geometry enabled by CTrex through operator‑level backprojection control. 
(\textbf{Left}) Conventional acquisition with the object fully covered by the detector field of view. 
(\textbf{Right}) eFOV acquisition with an offset centre of rotation, showing the central FOV region that is fully illuminated for all projection angles.  In eFOV CT, an offset centre of rotation increases the effective horizontal field of view, creating a central region fully illuminated for all projection angles and giving rise to eFOV‑specific artefacts addressed by modified backprojection kernels.}

    \label{fig:efov_schematic}
\end{figure}

\subsection{Extended fields of view}\label{sec:EFOV}

\begin{figure}
    \begin{subfigure}[c]{0.329\textwidth}
    \includegraphics[width=0.9\textwidth, height = 6cm]{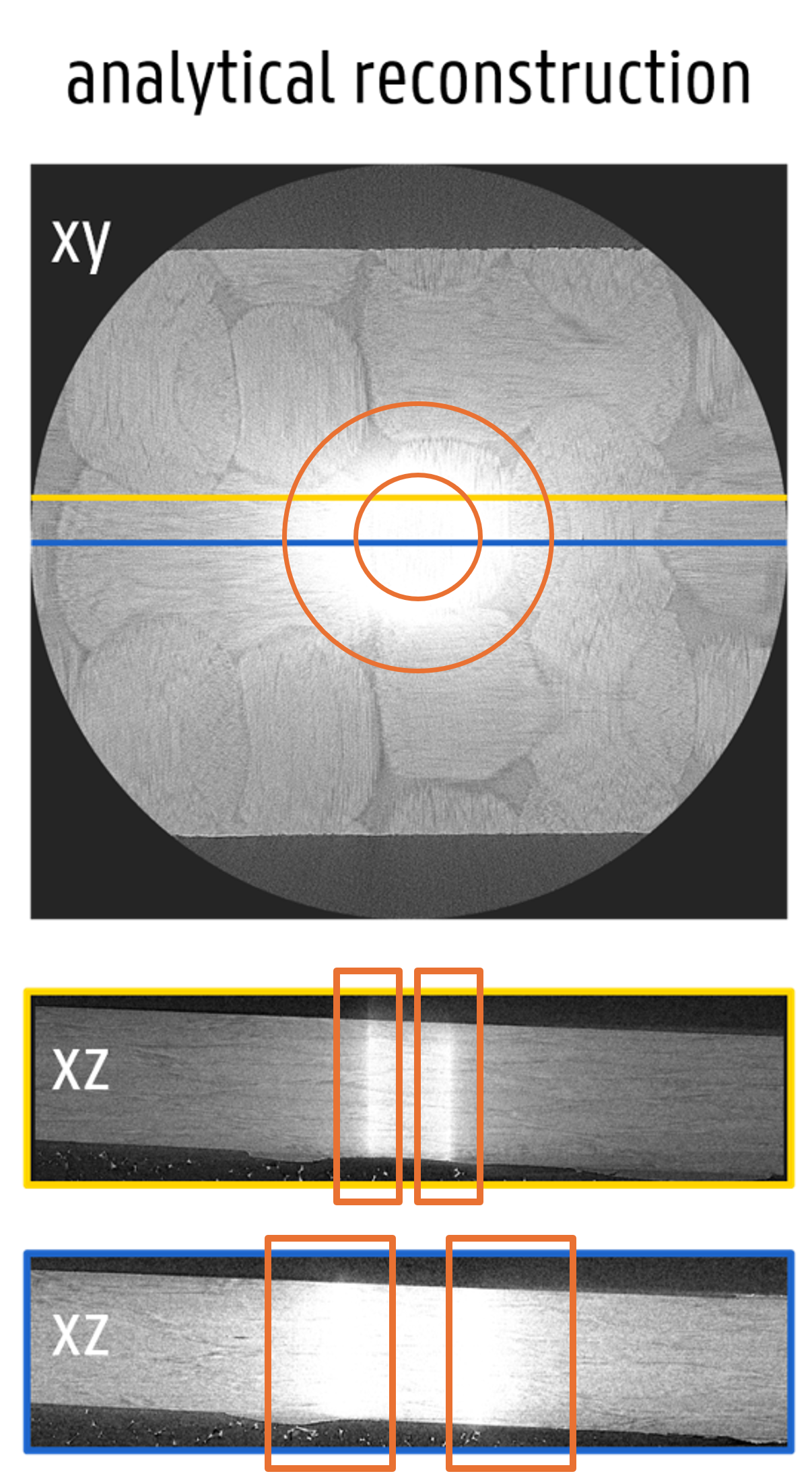}
    \captionsetup{width=.9\textwidth}
    \caption{Analytical reconstruction of a CBCT scan with offset COR leads to a central ring artefact if no proper weights are used to address the variation in projection coverage}
    \label{subfig:exp_eFOV_fdk}
    \end{subfigure}
    \begin{subfigure}[c]{0.329\textwidth}
        \includegraphics[width=0.9\textwidth, height = 6cm]{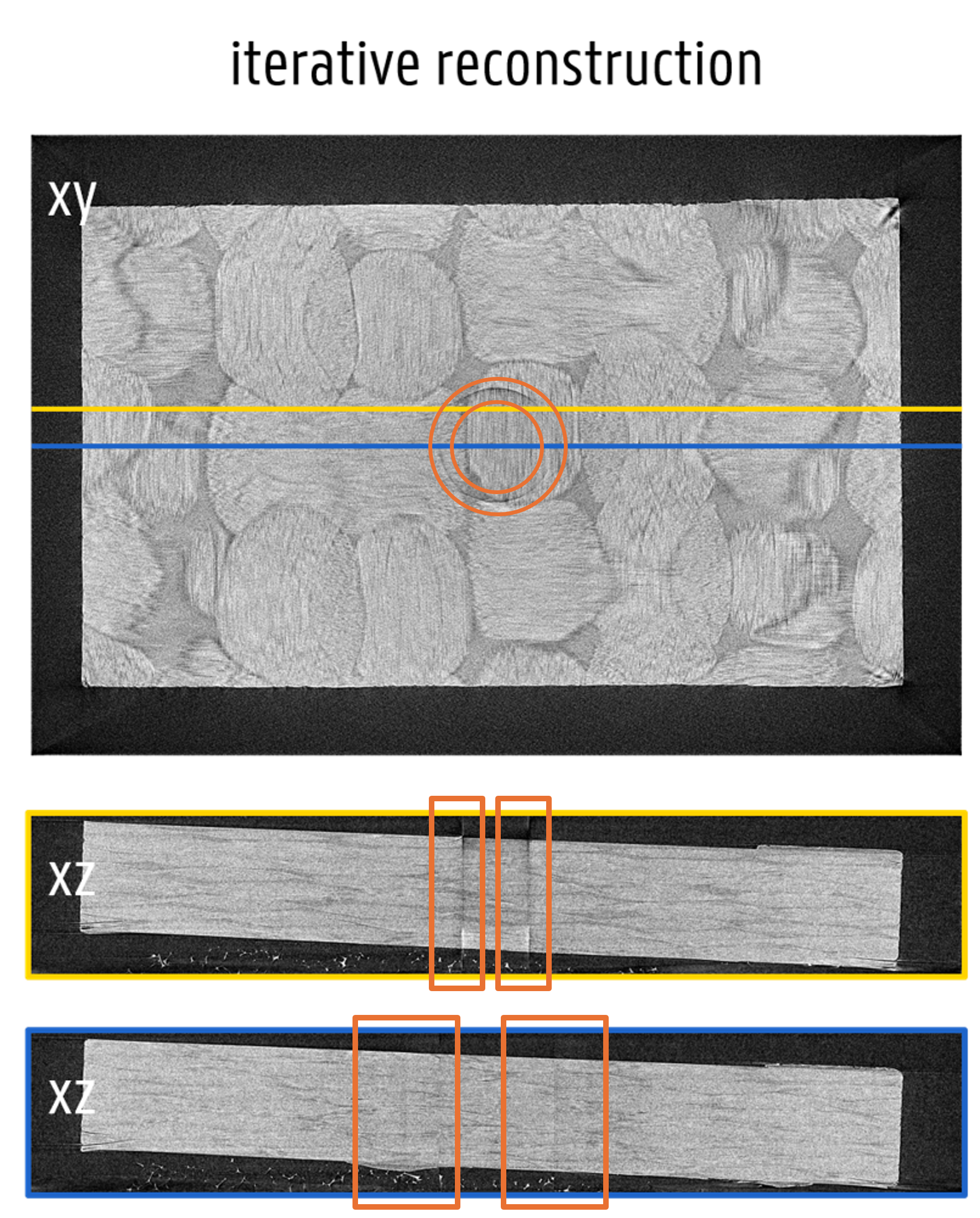}
        \captionsetup{width=.9\textwidth}
        \caption{Iterative reconstruction implicitly solves the weighting issue
so the simulated projections of this volume optimally match the acquired projections. Still,
a ring artefact is observed.}
        \label{subfig:exp_eFOV_norm}
    \end{subfigure}
    \begin{subfigure}[c]{0.329\textwidth}
        \vspace{-18pt}
        \includegraphics[width=0.9\textwidth, height = 6.2cm]{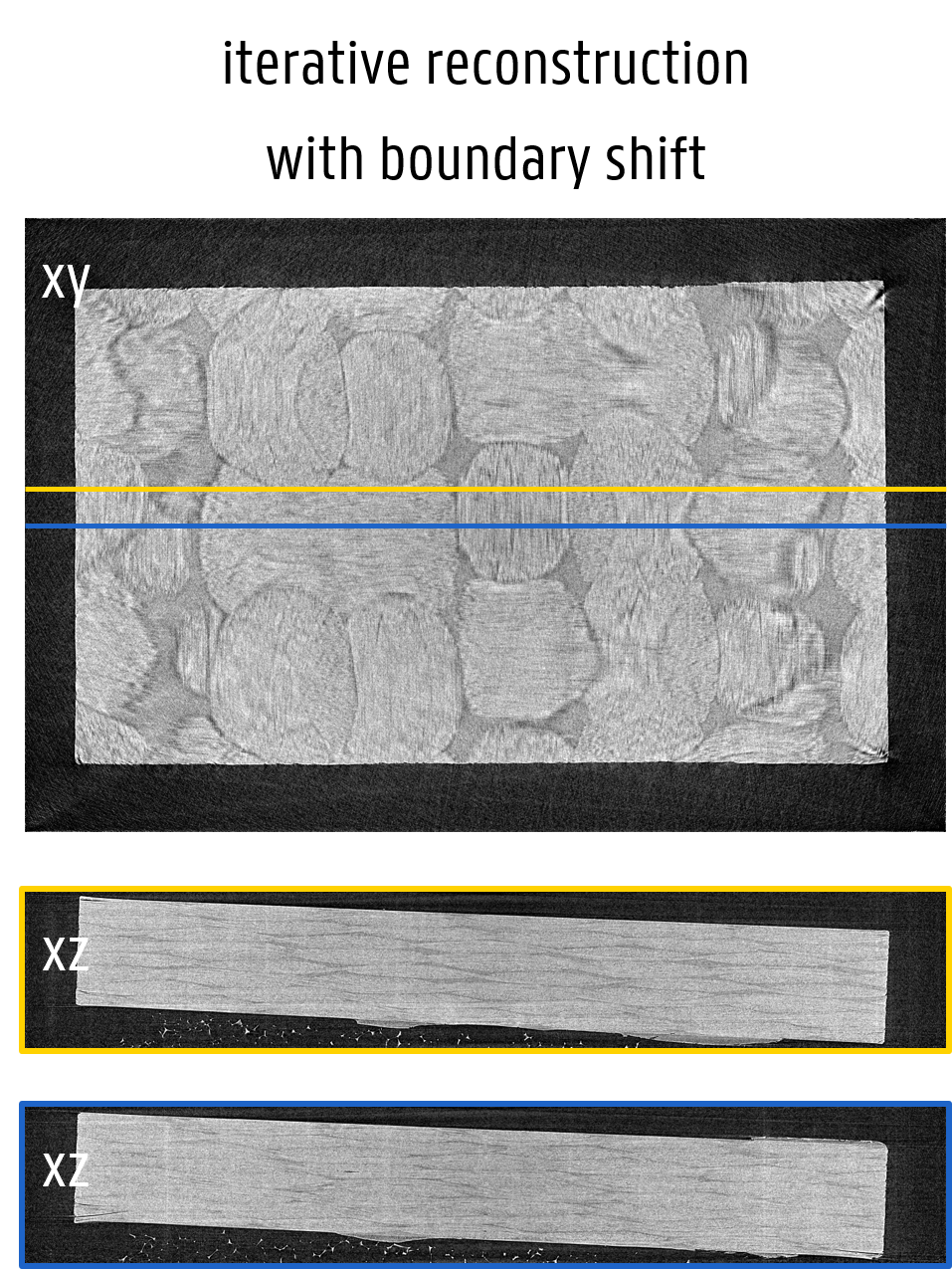}
        \captionsetup{width=.9\textwidth}
        \vspace{4pt}
        \caption{The iterative reconstruction of a scan with offset centre of rotation contains
no ring artefact when the boundary of the backprojection is shifted randomly.}
        \label{subfig:exp_eFOV_random}
    \end{subfigure}
     \caption{CTrex enabled extended field‑of‑view CBCT reconstruction of a carbon‑fibre reinforced polymer sample with offset centre of rotation \citep{thesis_wannes}. In Orange we indicated the vissable effect of reconstruction bias at the central FOV ring. 
(\subref{subfig:exp_eFOV_fdk}) Analytical reconstruction exhibiting a central FOV artefact. 
(\subref{subfig:exp_eFOV_norm}) Iterative reconstruction reducing central bias but showing a boundary artefact. 
(\subref{subfig:exp_eFOV_random}) Iterative reconstruction with modified backprojection kernel, suppressing the boundary artefact.}
    \label{fig:exp_eFOV}
\end{figure}

Extended field‑of‑view (eFOV) acquisition is commonly employed in industrial and materials‑science micro‑CT to image objects whose transverse extent exceeds the nominal detector width. In cone‑beam CT, the reconstructable object size is primarily limited by the horizontal field of view (FOV) of the detector, which for a central centre of rotation (COR) approximately equals the detector width. Offset centre‑of‑rotation scanning provides a practical means to increase the effective horizontal FOV, approaching up to twice the detector width in the ideal parallel‑beam limit. Such geometries are therefore frequently used to image oversized samples without reducing magnification or repositioning the object (Fig.~\ref{fig:efov_schematic}), but inherently introduce angle‑dependent truncation outside the fully illuminated central region ~\cite[Section 2.5.4]{thesis_wannes}.
\\\\
In these extended field‑of‑view (eFOV) acquisitions, only a central region of the object remains within the detector FOV for all projection angles, while outer regions fall outside the detector FOV for part of the acquisition. As a result, measurements of these regions are incomplete and vary with projection angle. Although the acquisition geometry satisfies the Tuy–Smith conditions \citep{tuyInversionFormulaConeBeam1983, smithImageReconstructionConeBeam1985}, and is therefore sufficient for reconstruction in principle, this incomplete angular coverage limits how reliably features outside the fully illuminated region can be recovered. This effect is clearly visible in analytical reconstruction, where truncation causes strong over‑attenuation artefacts in the central FOV region (Fig.~\ref{subfig:exp_eFOV_fdk}). Iterative reconstruction reduces these large‑scale effects, but a persistent boundary artefact remains at the interface between the fully illuminated and partially observed regions, referred to as the central FOV ring (Fig.~\ref{subfig:exp_eFOV_norm}). At this boundary, voxels near the edge of the detector support are repeatedly updated using measurements that partially integrate attenuation across the truncation boundary. Because the effective support of these measurements changes with projection angle, a systematic bias is introduced and reinforced over successive iterations, producing a staircase‑like artefact at the central FOV boundary (indicated in orange border in Fig.~\ref{subfig:exp_eFOV_norm}).
\\\\
Within CTrex, this effect can be mitigated by introducing controlled variation in the backprojection boundary. During backprojection, the effective detector edge is shifted per projection by a small random offset between the physical detector boundary and the projected centre of rotation. While the forward projection continues to evaluate the full detector extent, this prevents voxels near the truncation boundary from repeatedly encountering the same sampling configuration and ensures that they also receive updates from projections in which they are not systematically aligned with the detector edge. Over successive iterations, this variation disrupts the accumulation of boundary‑induced bias, resulting in a reduced staircase artefact at the central FOV ring (Fig.~\ref{subfig:exp_eFOV_random}).

\subsection{Multi‑scale imaging via ROI reconstruction}\label{sec:ROI}

In many materials imaging applications, fine‑scale structures critically influence macroscopic behaviour, while a sufficiently large field of view is required to place these features in their structural context. Achieving high spatial resolution without sacrificing global coverage therefore motivates multi‑scale imaging strategies based on region‑of‑interest (ROI) acquisition.
\\\\
In ROI scanning, a small region of the sample is imaged at higher magnification within a larger object that is only sparsely or partially sampled at that resolution. As a result, the measured projections contain contributions from material outside the high‑resolution ROI that cannot be directly reconstructed at the same scale, which leads to strong artefacts if not properly accounted for.
\\\\
Iterative reconstruction provides a natural framework for addressing this challenge by explicitly modelling the contribution of the surrounding volume during reconstruction. One practical approach is to first reconstruct the entire object on a coarse grid, capturing the dominant attenuation contributions of the full sample. These contributions are then forward projected and subtracted from the measured data, yielding corrected projections that predominantly contain information from the ROI. The ROI can subsequently be reconstructed at higher resolution using the corrected data.
\\\\
An alternative strategy reconstructs the full object at low resolution and embeds a higher‑resolution ROI directly within the same reconstruction process. Although this introduces coupling artefacts at the ROI boundary, these can be mitigated by reconstructing a slightly expanded ROI and discarding the border region after convergence.
\\\\
The multi‑resolution approach introduced by De Witte \textit{et al.}~\citep{multi_res} generalizes these ideas by combining coarse global reconstruction with multiple adjacent high‑resolution ROI reconstructions. In CTrex, this formulation is implemented by adapting the projection kernels to simultaneously handle multiple resolution levels within a single forward and backward projection pass (Sec.~\ref{sec:P_kern_ROI}). This kernel‑level flexibility enables efficient multi‑scale reconstruction without restructuring the iterative control logic.
In conventional frameworks, such multi‑resolution coupling typically requires separate reconstructions, iterative subtraction of coarse‑scale projections, or repeated forward projections of intermediate volumes. These approaches introduce additional full‑volume passes and memory overhead, making tightly coupled multi‑scale reconstruction difficult or inefficient.

\begin{figure}
    \begin{subfigure}[c]{0.45\textwidth}
    \includegraphics[width=0.95\textwidth]{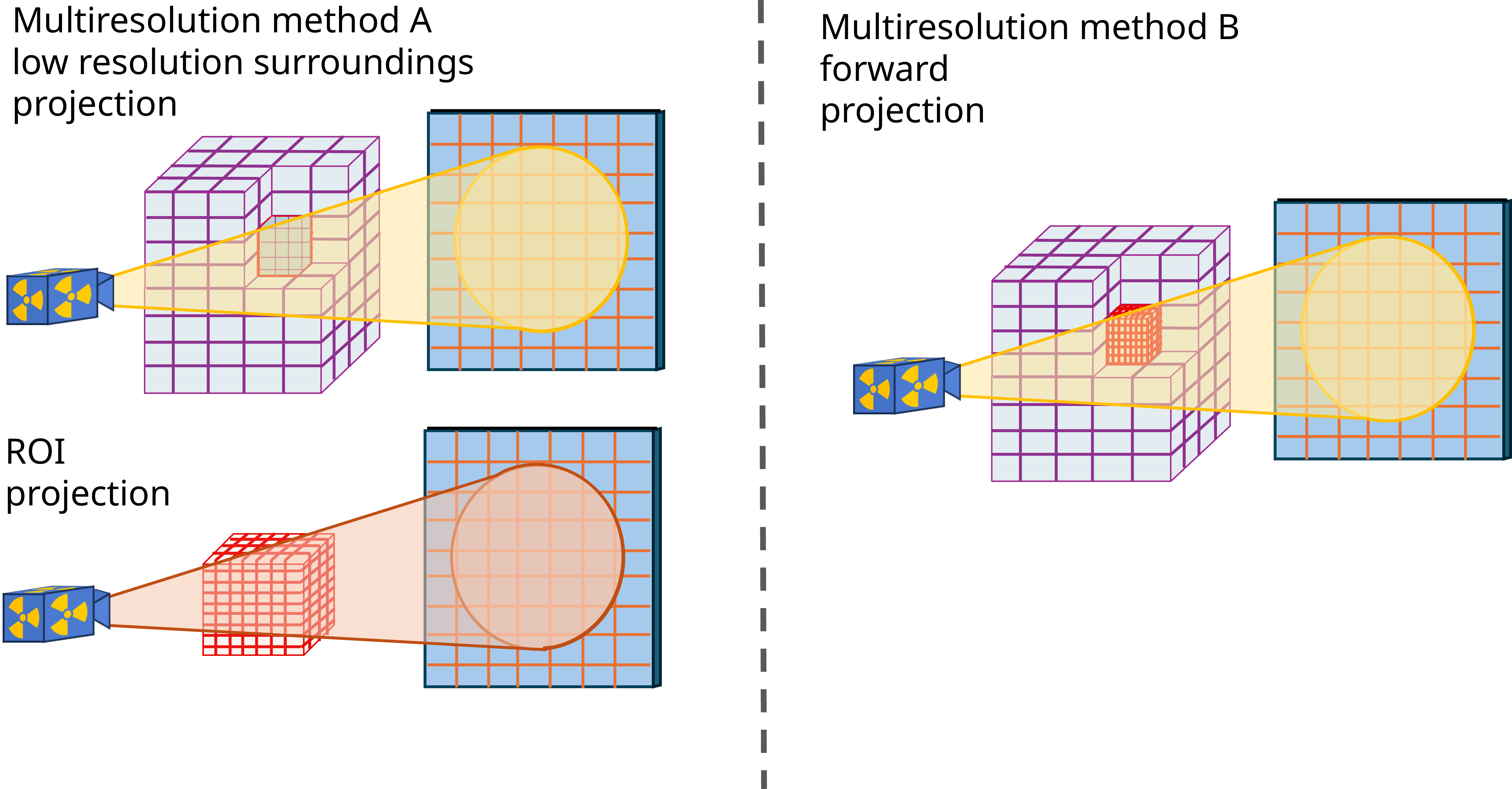}
    \captionsetup{width=.9\textwidth}
    \caption{\emph{Left Method A} Illustration of two step approach. first low resolution projection simulation of only surrounding volume (see hollow cube).
    \emph{Right method B} Illustration of simultaneous projection of low and high resolution volumes.
    }
    \label{fig:ROI-explain}
    \end{subfigure}
    \begin{subfigure}[c]{0.45\textwidth}
    \includegraphics[width=0.95\textwidth]{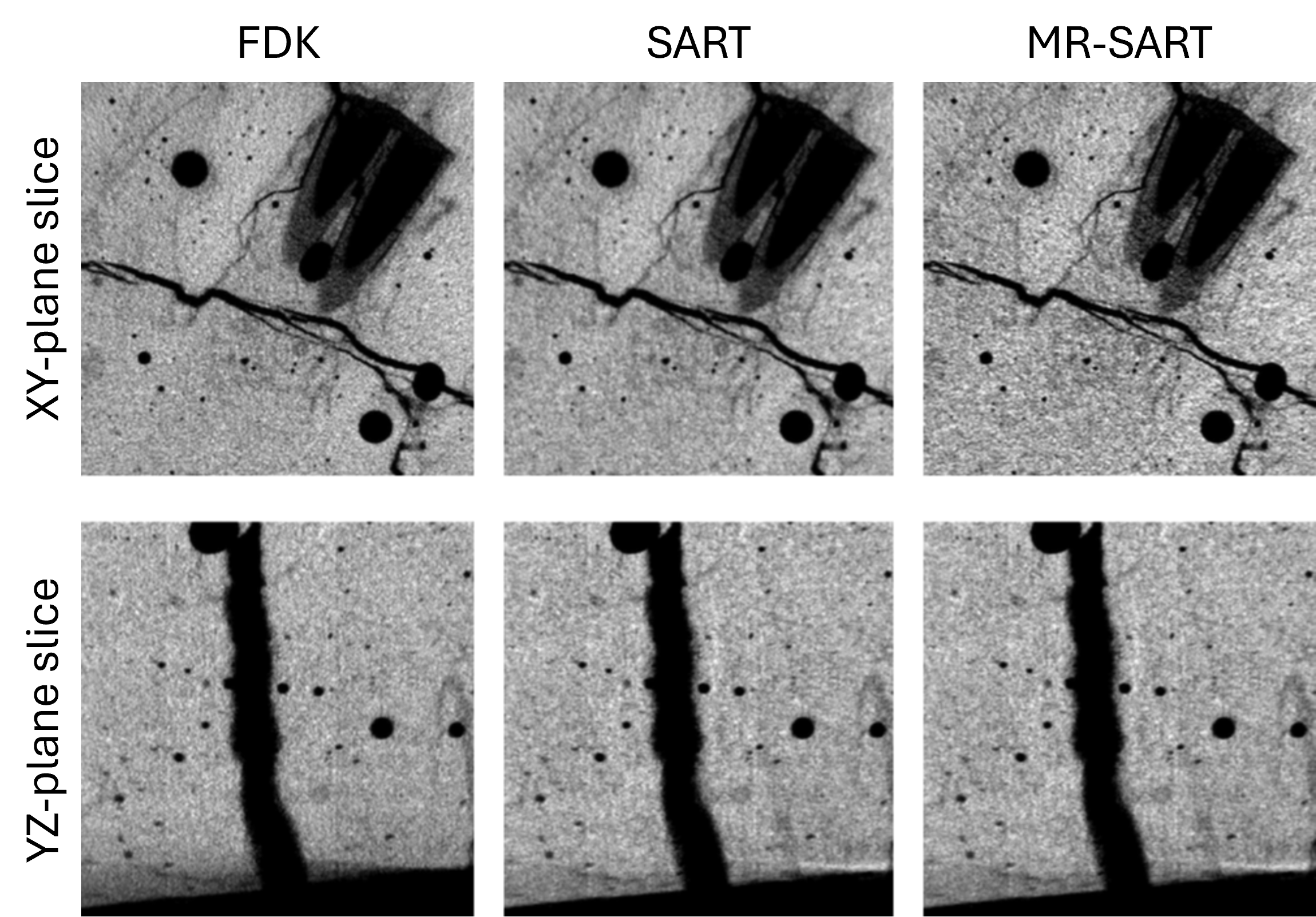}
    \captionsetup{width=.9\textwidth}
    \caption{reconstructed cross sections of a piece of self-healing cement, using FDK (left), SART (middle) and MR-SART [Multi Resolution SART] (right). With upper row XY-plane slice at z =-0.25 and lower row the YZ-plane slice. \citep{multi_res}}
    \label{fig:ROI-example}
    \end{subfigure}
    \caption{Multi‑scale ROI reconstruction.
(\textbf{Left}) Schematic illustration of two multi‑resolution projection strategies: subtraction‑based ROI reconstruction and simultaneous multi‑resolution projection. 
(\textbf{Right}) Cross‑sections of a self‑healing cement sample reconstructed using FDK, SART, and multi‑resolution SART (MR‑SART), showing improved local reconstruction quality in the ROI \citep{multi_res}.}
    \label{fig:ROI}
\end{figure}

\section{Technical description of kernel‑level adaptations}
\label{sec:kernel_level_adap}

The application studies in Section~\ref{sec:application_study} show that several practical reconstruction challenges originate from how the forward and backward projection operators are evaluated, rather than from the high‑level structure of the reconstruction algorithm itself. Addressing such effects therefore requires targeted modifications at the level of the projection and backprojection kernels, instead of adjustments limited to iteration parameters or regularization strategies.
\\\\
This section provides a technical description of the kernel‑level adaptations that were implemented in CTrex to make the presented application case studies possible. The aim is not to introduce new reconstruction algorithms, but to document how deliberate modifications of the projection, correction, and backprojection operators enable the reconstruction strategies discussed earlier, while preserving the overall iterative control structure managed by the \texttt{Rex} class. For clarity, we first provide a brief refresher on the baseline implementation of the ordered‑subset SIRT projection, correction, and backprojection operators, after which the kernel‑level adaptations are described. The section is organized by operator type, addressing projector kernels first and then backprojector kernels.

\subsection{Baseline reconstruction operators}\label{sec:kernels}

As described in Sec.~\ref{sec:iterative_schemes}, the \texttt{Rex} class controls the ordering and scheduling of reconstruction operators within each iteration of the ordered‑subset SIRT update scheme. The numerical operators appearing in Eq.~\eqref{eq:sirt} namely the forward projection $A$ [\textbf{1}], the detector‑space correction operator $R(\cdot)$ [\textbf{2}], and the backprojection with voxel‑space normalization $C A^{T}$ [\textbf{3}] are realized as three GPU kernels.
\\\\
The baseline kernel implementations define the standard projection, correction, and backprojection operators, while reconstruction variants such as SIRT, SART, and ordered‑subset schemes are obtained through the scheduling logic implemented by \texttt{Rex}. In CTrex, these kernels can subsequently be adapted in a controlled manner when problem‑specific modifications of the operator evaluation are required, as described in the remainder of this section.

\begin{figure}
    \centering
    \captionsetup[subfigure]{aboveskip=10pt} 

    \begin{subfigure}[t]{0.32\textwidth}
        \centering
        \vbox to 5cm{\vfill \includegraphics[height=4.5cm]{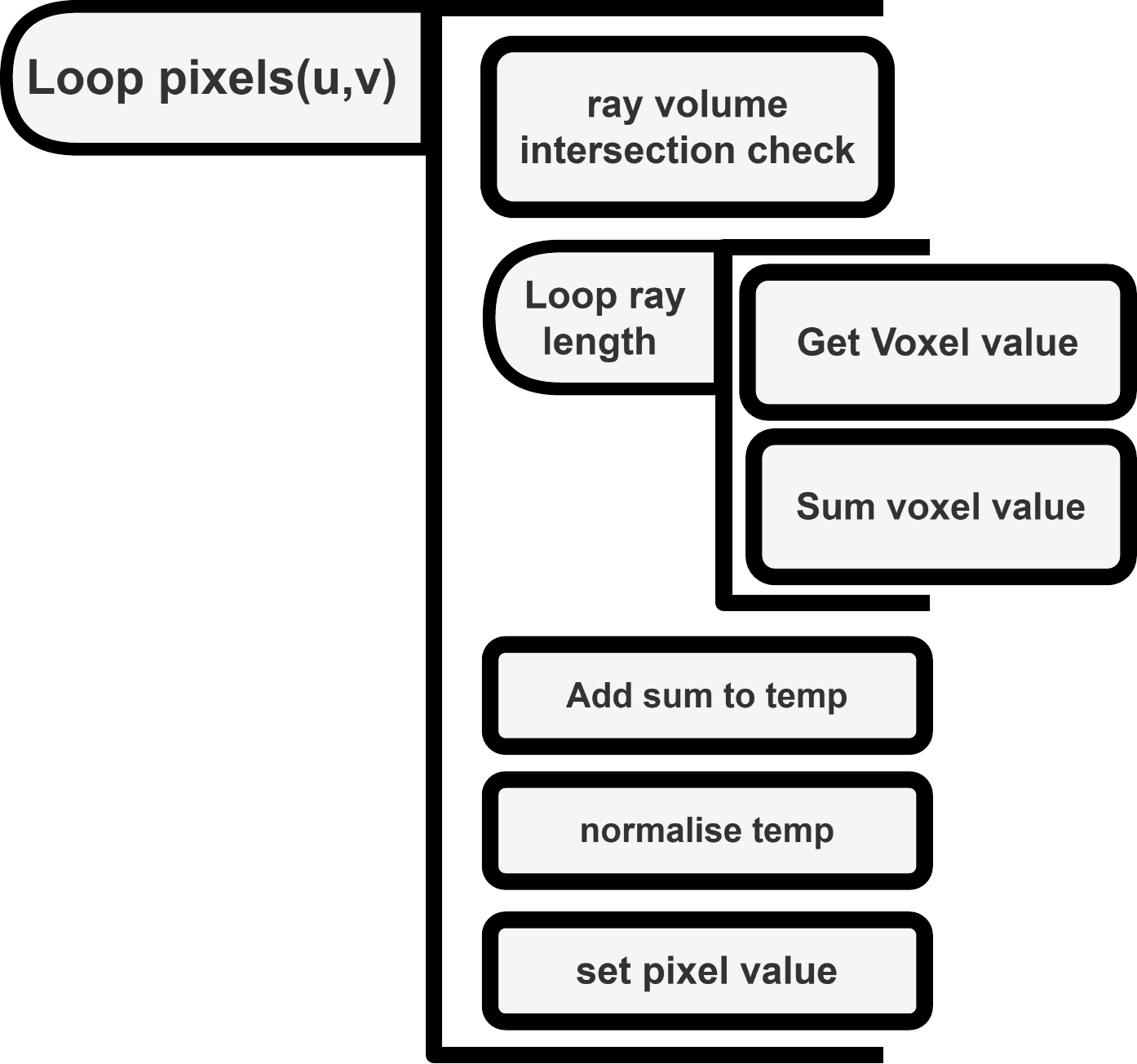} \vfill}
        \caption{Block diagram of projection kernel. The kernel implementation of operator $A\mu_k$, annotated as \textbf{1} in Eq. \ref{eq:sirt}}
        \label{subfig:proj_code}
    \end{subfigure}
    \hfill
    \begin{subfigure}[t]{0.32\textwidth}
        \centering
        \vbox to 5cm{\vfill \includegraphics[height=4.5cm]{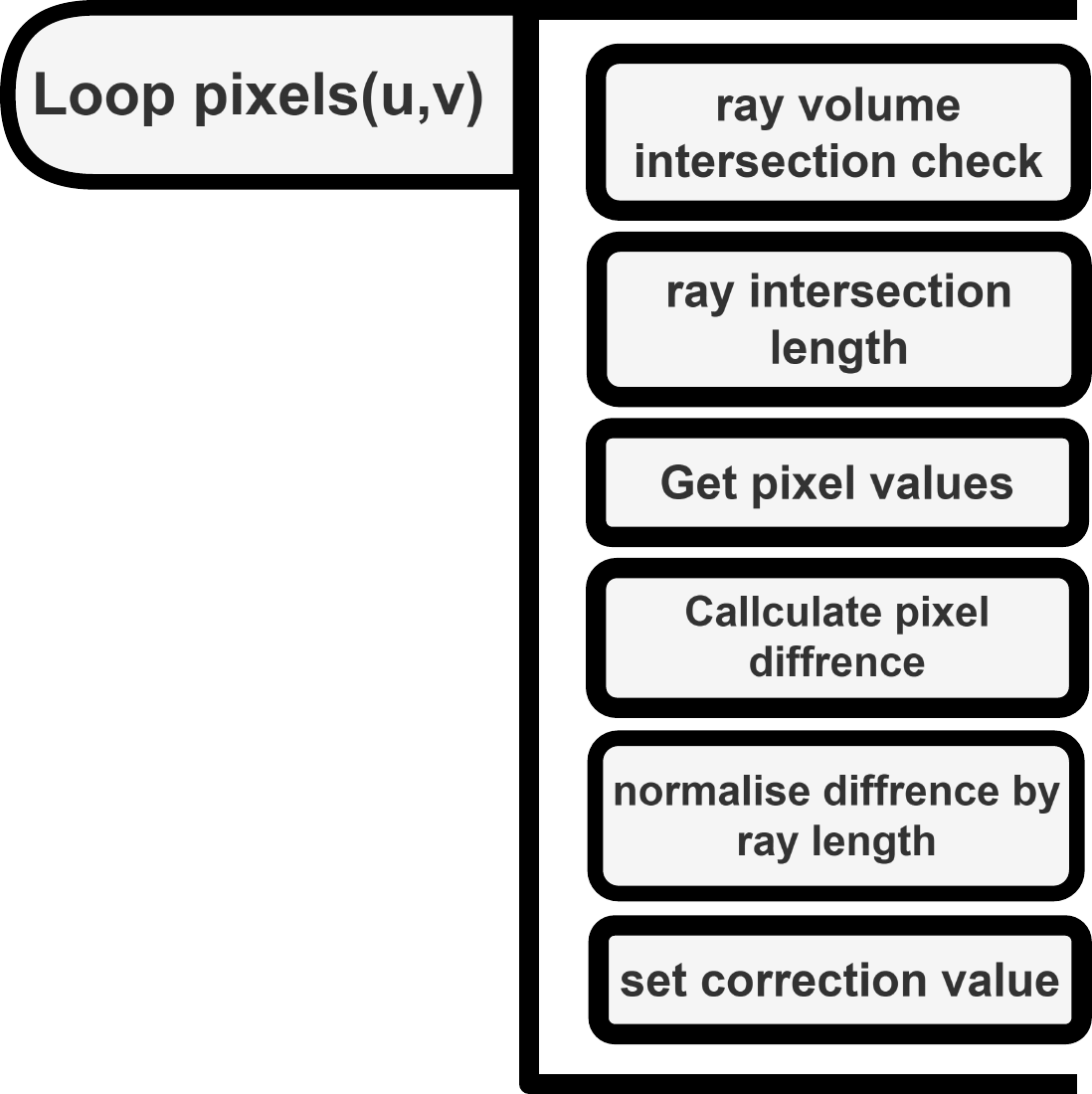} \vfill}
        \caption{Block diagram of correction kernel. The kernel implementation of operator $R\!\left(p - A\mu_k\right)$ annotated as \textbf{2} in Eq. \ref{eq:sirt}}
        \label{subfig:corr_code}
    \end{subfigure}
    \hfill
    \begin{subfigure}[t]{0.32\textwidth}
        \centering
        \vbox to 5cm{\vfill \includegraphics[height=4.5cm]{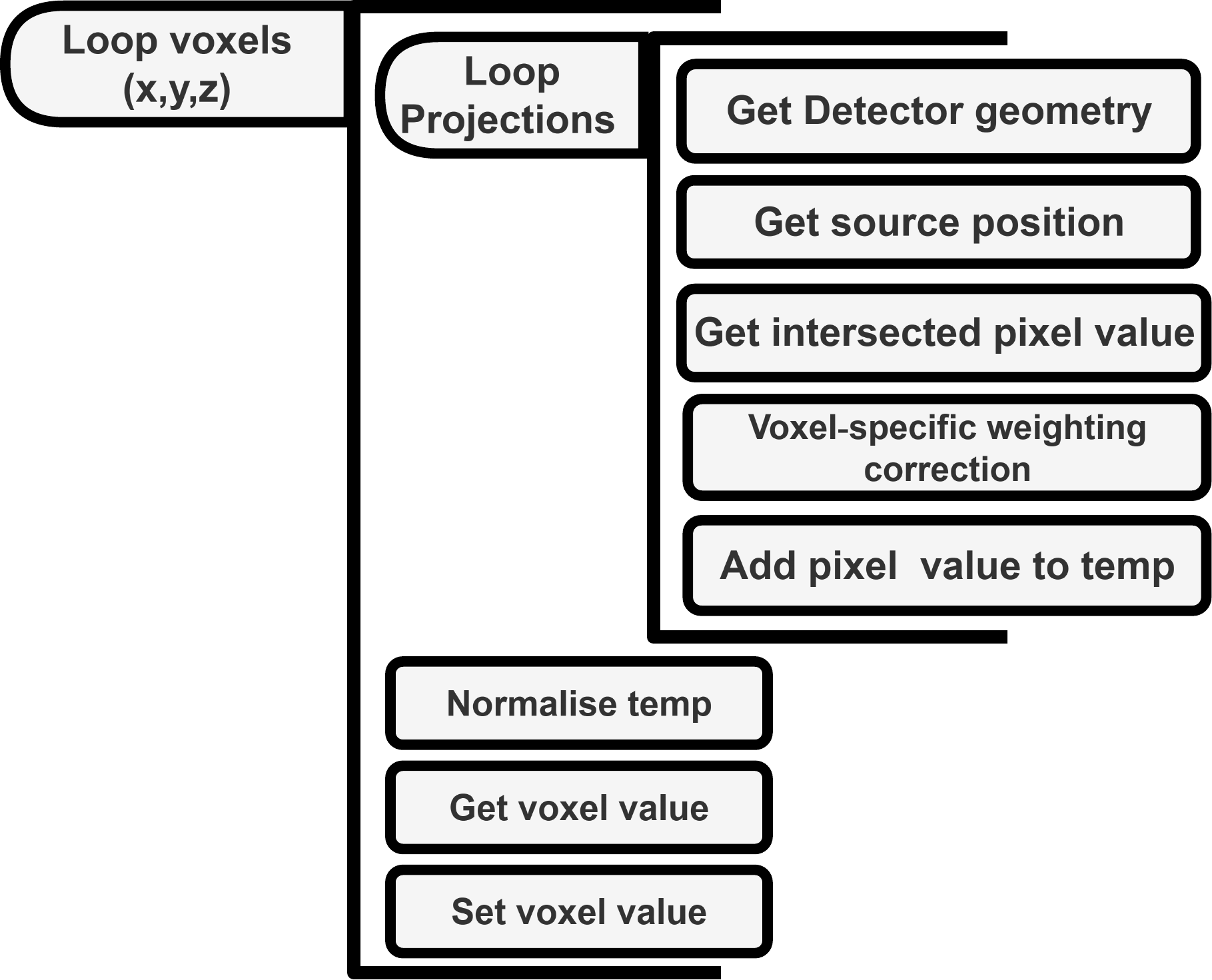} \vfill}
        \caption{Block diagram of backprojection kernel. The kernel implementation of operator $CA^{T}$ thus indicated as step \textbf{3} in Eq. \ref{eq:sirt}}
        \label{subfig:backproj_code}
    \end{subfigure}
    
    \caption{The block diagrams of the projection, correction, and backprojection kernels.}
    \label{fig:corect_and_backproject}
\end{figure}

\paragraph{Projection operator ($A\mu_k$)}\label{sec:proj_kernel}

The projection kernel implements the forward projection operation $A\mu_k$ [annotated as \textbf{1} in Eq.~\eqref{eq:sirt}]. For each detector pixel, a ray is constructed from the X‑ray source to the pixel centre using the per‑projection geometry supplied by the \texttt{Trajectory} abstraction. All rigid or affine transformations of the source, detector, and volume are applied prior to ray traversal so that sampling is performed entirely in the volume‑attached coordinate system.
\\\\
The ray is intersected with the reconstruction volume, and the segment inside the volume is sampled using a pixel‑driven implementation of Joseph’s method \citep{josephFastRayTracing1982}. Attenuation values are obtained via hardware‑assisted trilinear interpolation of the volume texture. The samples are accumulated and weighted by their corresponding path lengths to produce a simulated detector value consistent with the measured projection data (Fig.~\ref{subfig:proj_code}).

\paragraph{Correction kernel}\label{sec:corr_kernel}

The correction kernel corresponds to the residual and correction‑space operations in Eq.~\eqref{eq:sirt}, i.e., 
$R\!\left(p - A\mu_k\right)$,
annotated as \textbf{2}. For each detector pixel, the kernel subtracts the simulated intensity $A\mu_k$ from the measured intensity $p$, yielding a \emph{correction term in projection (detector) space} that quantifies the mismatch between prediction and measurement. These correction terms represent the quantities that must be propagated back into the reconstruction volume in the subsequent backprojection step.
\\\\
This operation is performed entirely in detector space and is efficiently implemented using texture memory for accessing the simulated projections. After computing the per‑pixel residual, the correction term may optionally be weighted or normalized, for example to apply geometric weighting, statistical weighting, or ordered‑subset selection as defined by the \texttt{Rex} class. Because the correction kernel provides a clean separation between forward projection and backprojection, it enables the incorporation of algorithmic modifications such as dynamic weighting or region‑specific scaling without requiring changes to the projector or backprojector implementations themselves (Fig.~\ref{subfig:corr_code}).

\paragraph{Backprojection operator ($C A^{T}$)}\label{sec:backproj_kernel}

The backprojection kernel implements the transpose operator $A^{T}$ together with voxel‑space normalization $C$, corresponding to the update step annotated as \textbf{3} in Eq.~\eqref{eq:sirt}. For each voxel and for each correction image selected by the \texttt{Rex} iteration logic, a ray is constructed from the source to the voxel centre. The corresponding detector coordinates are computed, and the detector‑space correction image is sampled to obtain the contribution for that voxel.
\\\\
The accumulated contributions are normalized to compensate for non‑uniform ray coverage, for example using precomputed path‑length factors or geometry‑dependent normalization weights. The resulting update is scaled by the relaxation parameter $\alpha$ and applied to produce the next volume estimate $\mu_{k+1}$. A schematic overview of the baseline backprojection kernel is shown in Fig.~\ref{subfig:backproj_code}.

\subsection{Projector kernel adaptations}

The projector kernel defines how rays are generated, how they traverse the reconstruction domain, and how attenuation values are sampled and integrated. Modifying this kernel therefore enables changes at several conceptual levels: the geometric definition of the volume, the sampling locations along each ray, and the numerical integration of contributions into detector pixels. The adaptations described below illustrate how CTrex allows intervention at each of these levels while preserving the overall ray‑driven structure of the baseline projector (Sec.~\ref{sec:proj_kernel}). A schematic overview of the projector‑kernel adaptations discussed in the following subsections is provided in Fig.~\ref{fig:project-mods}.

\subsubsection{Geometry‑level modifications}

\begin{figure}
    \centering
    \includegraphics[width=0.75\textwidth]{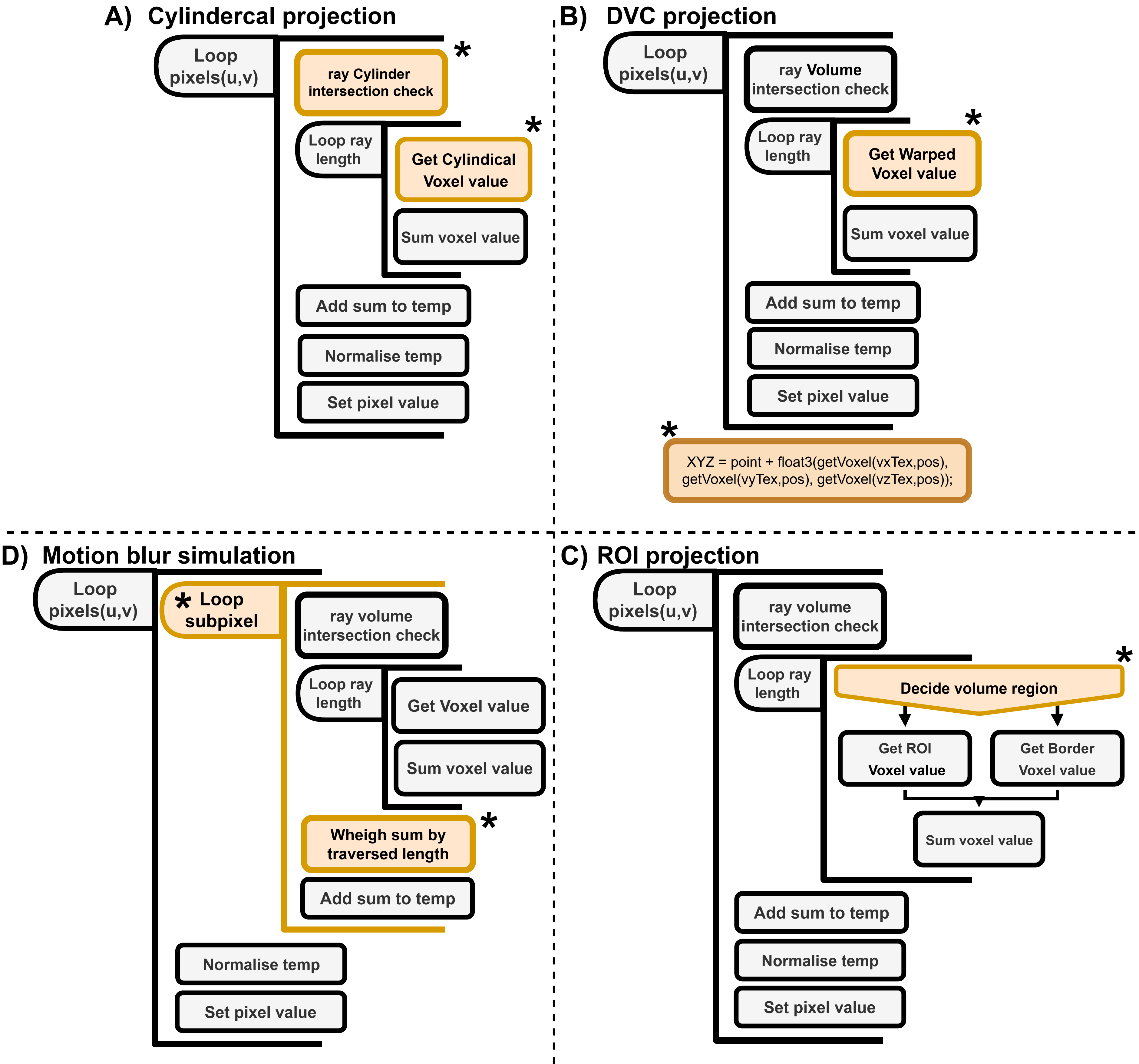}
    \caption{Schematic overview of projector kernel adaptations: geometry‑level modifications (cylindrical domain, DVC deformation) and sampling‑level modifications (motion‑blur integration, ROI sampling). Modified code blocks relative to the baseline projector kernel (Fig.~\ref{subfig:proj_code}) are indicated by $\boldsymbol{*}$.}
    \label{fig:project-mods}
    \phantomsubcaption\label{subfig:cylinder-proj}
    \phantomsubcaption\label{subfig:dvc-proj}
    \phantomsubcaption\label{subfig:motion-blur}
    \phantomsubcaption\label{subfig:roi-project}
    
\end{figure}

The most fundamental level of projector adaptation concerns how the reconstruction domain itself is represented and intersected by rays. In these cases, the ray generation logic remains unchanged, but the mapping between ray coordinates and volume coordinates is modified.
\\\\
In the cylindrical reconstruction case (Sec.~\ref{sec:cylinder}), the standard ray bounding‐box intersection test is replaced by an intersection with a bounding cylinder (Fig.~\ref{subfig:cylinder-proj}). Ray traversal proceeds only if such an intersection exists. Sample points along the ray are then mapped from Cartesian to cylindrical coordinates prior to volume sampling, and the reconstruction volume is stored on an $N_r \times N_\phi \times N_z$ grid compatible with GPU texture memory and trilinear interpolation.\label{sec:P_kern_cyl} Apart from this coordinate transformation, the ray sampling and integration logic remains identical to the baseline projector.
\\\\
To accommodate misalignment between the rotation axis and the reconstruction domain, additional rigid transformations are applied during the coordinate mapping step. These transformations allow the cylindrical volume to have an independent pose with respect to the laboratory reference frame. While such transformations can in principle be estimated as part of the reconstruction process, only explicit specification was considered here.
\\\\
A related geometry‑level modification is used in DVC‑based deformation-aware projection (Sec.~\ref{sec:DVC}). In this case, a non‑parametric displacement field $\mathbf{u}(\mathbf{x})$ is defined over the reconstruction volume. During ray traversal, the displacement field is interpolated at each sample position and used to warp sampling coordinates prior to volume lookup. This effectively modifies the volume geometry locally without altering the ray geometry itself (Fig.~\ref{subfig:dvc-proj}).\label{sec:P_kern_DVC} The additional cost is limited to a small number of texture fetches per sample.
\subsubsection{Sampling and loop‑level modifications}

Beyond geometric mapping, projector behaviour can be altered by modifying the internal loop structure that performs ray sampling and integration. These adaptations change how detector values are accumulated from multiple contributions along a ray.
\\\\
For motion‑blur modelling (Sec.~\ref{sec:motion}), an additional integration loop is introduced inside the per‑projection loop.\label{sec:P_kern_motion} Each detector pixel is subdivided into multiple angular or temporal sub‑samples, and a separate ray traversal is performed for each sub‑sample. The resulting contributions are accumulated to approximate the continuous integration performed during fly‑scan acquisition (Fig.~\ref{subfig:motion-blur}). Importantly, the correction and backprojection operators remain unchanged; only the projector encodes the motion‑aware integration.
\\\\
In multi‑resolution ROI reconstruction (Sec.~\ref{sec:ROI}), the sampling logic is extended to support traversal of multiple volumes at different resolutions.\label{sec:P_kern_ROI} Rays may pass through both a high‑resolution ROI volume and a lower‑resolution surrounding context volume, with sampling switched dynamically based on ray position (Fig.~\ref{subfig:roi-project}). This modification affects only the sampling stage of the projector and requires no changes to ray generation or detector accumulation logic.

\subsection{Backprojector kernel adaptations}

The backprojector kernel determines how detector‑space correction values are accumulated into the reconstruction volume. Kernel‑level modifications at this stage can affect how updates are spatially distributed, weighted, or normalized, and are therefore particularly relevant in the presence of motion, incomplete data, or non‑standard reconstruction domains. The following subsections describe representative backprojector adaptations used in the application studies, with an overview of the modified accumulation patterns shown in Fig.~\ref{fig:back-mods}.

\subsubsection{Geometry‑level modifications}

\begin{figure}
    \centering
    \includegraphics[width=0.75\textwidth]{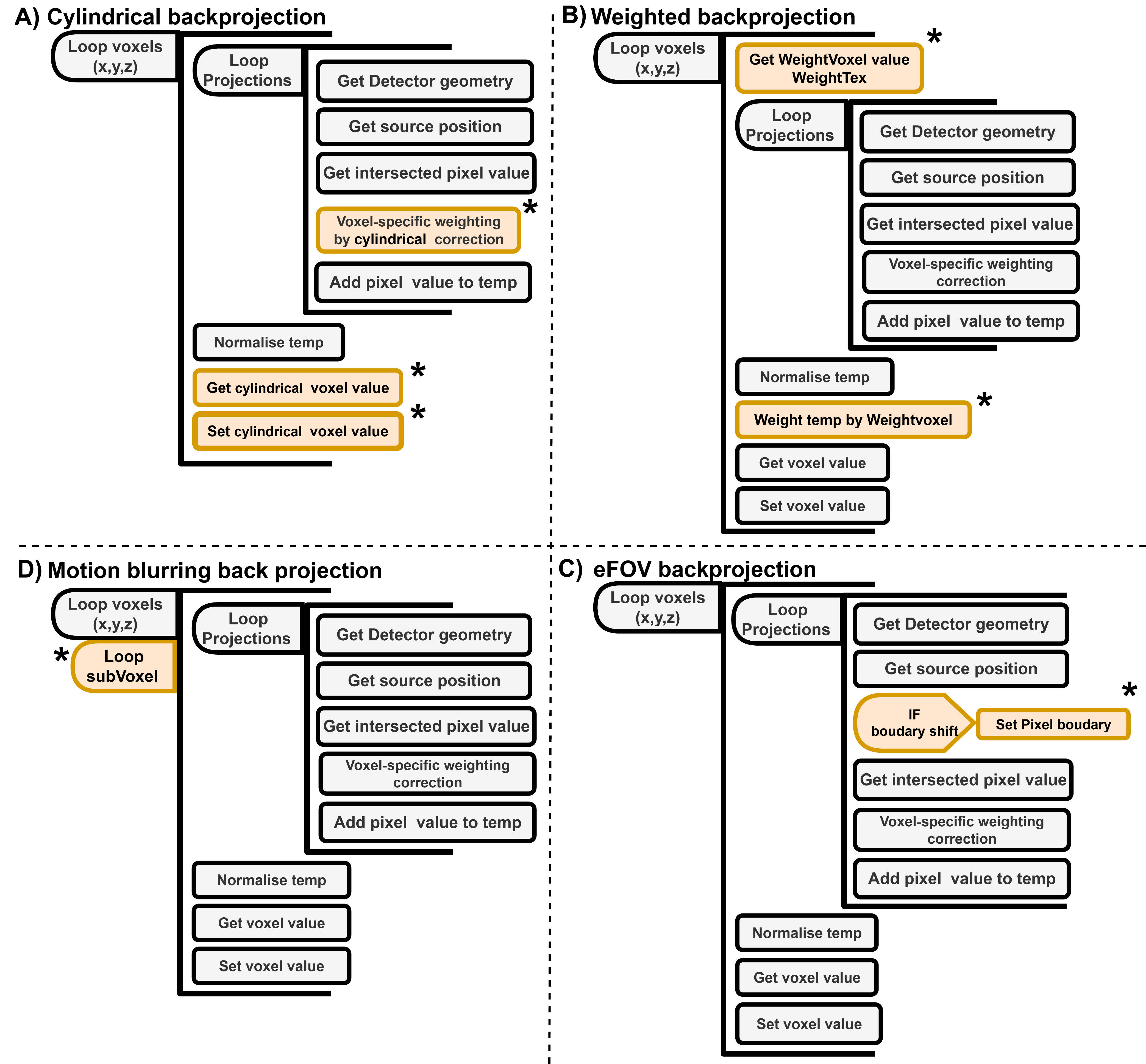} 
    \caption{Schematic overview of backprojector kernel adaptations, grouped by geometry‑level (cylindrical domain, DVC deformation) and accumulation‑level modifications (motion‑blur integration, eFOV sampling). Modified code blocks relative to the baseline backprojection kernel (Fig.~\ref{subfig:backproj_code}) are indicated by $\boldsymbol{*}$.}
    \label{fig:back-mods}
    \phantomsubcaption\label{subfig:cylindrical-backproj}
    \phantomsubcaption\label{subfig:motion-blur-back}
    \phantomsubcaption\label{subfig:efov-back}
    \phantomsubcaption\label{subfig:weighted-back}
    
\end{figure}

Geometry adaptations in the backprojector mirror those applied in the projector to ensure consistency between forward and inverse operators.\label{sec:B_kern_cyl} In the cylindrical reconstruction case (Sec.~\ref{sec:cylinder}), voxel iteration is performed in cylindrical coordinates, while detector coordinates are obtained by mapping voxel positions back into Cartesian space prior to detector lookup (Fig.~\ref{subfig:cylindrical-backproj}). Apart from this coordinate transformation and the associated volume layout, the accumulation logic is unchanged.

\subsubsection{Accumulation and weighting modifications}

More substantial behavioural changes arise when modifying how correction values are accumulated into the volume.
Weighted backprojection (Sec.~\ref{sec:weight}) introduces voxel‑wise weights $w_j$ that scale the contribution received by each voxel during backprojection.\label{sec:B_kern_weight} These weights are applied directly inside the accumulation loop (Fig.~\ref{subfig:weighted-back}), allowing dynamic regions to absorb a larger fraction of projection‑space corrections while static regions evolve more conservatively.
\\\\
For motion‑blur compensation (\ref{sec:motion}),\label{sec:B_kern_motion} complementary modifications to the projection kernel can be applied at the level of the backprojector. While the projector integrates multiple sub‑samples per detector pixel, the backprojector accumulates the corresponding correction values onto a higher‑resolution voxel grid. This sub‑voxel backprojection does not introduce temporal sub‑sampling, but improves interpolation consistency with the motion‑aware projection operator and reduces discretization error in subsequent forward projections. The modification is confined to the accumulation loop and requires no changes to the ray geometry or correction operator (Fig.~\ref{subfig:motion-blur-back}). It comes at the cost of increased memory usage and computational load.
\\\\
Extended field‑of‑view reconstruction (Sec.~\ref{sec:EFOV}) illustrates another class of accumulation‑level intervention.\label{sec:B_kern_EFOV} At the boundary between fully and partially illuminated regions, a fixed detector boundary leads to systematic sampling bias and staircase artefacts. This is mitigated by introducing a small randomized shift in the effective detector boundary during backprojection (Fig.~\ref{subfig:efov-back}). Over successive iterations, this disperses boundary‑related bias into already‑converged regions without altering the global update scheme.

\section{Discussion}\label{sec:discussion}

CTrex, the framework presented in this work is designed as a research‑oriented infrastructure for developing and evaluating reconstruction methods in which image quality is strongly influenced by how the forward and backward operators are formulated and evaluated. Rather than functioning as a closed reconstruction pipeline, the framework exposes projection, correction, and backprojection operators as editable computational components while offering a unified iterative control structure. This makes it possible to investigate reconstruction problems in which the operator fidelity, rather than algorithmic scheduling alone, determines achievable image quality.
\\\\
A central outcome of the presented application studies is that operator‑level modifications can be implemented within a single framework without restructuring the surrounding reconstruction logic. In several cases, alternative implementations would require repeated full‑volume resampling, auxiliary reconstruction stages, or indirect preprocessing and post‑processing steps. These workarounds increase computational cost, complicate implementation, and make it more difficult to preserve a physically interpretable reconstruction model. By contrast, integrating such effects directly into the projection and backprojection operators allows them to be handled where they arise, namely at the level of image formation and reconstruction.
\\\\
From a methodological perspective, separating iteration control from operator implementation enables algorithmic exploration of modelling assumptions independently of the reconstruction scheme, enabling reconstruction research that is difficult to conduct within black‑box frameworks. It therefore becomes possible to explore how different modelling assumptions regarding motion, geometry, sampling density, or spatial prioritization affect convergence behaviour and reconstruction quality, while keeping the surrounding reconstruction logic unchanged. Moreover, this allows combinations of adaptations such as motion compensation, spatial weighting, and non‑standard geometry to be investigated within a consistent framework, without requiring combinatorial growth in algorithm complexity.
\\\\
A further consequence of this design is that non‑standard geometries and time‑dependent acquisition models can be incorporated without redesigning the reconstruction framework itself. The explicit geometric formulation based on view matrices allows scan trajectories, detector misalignment, and affine transformations to be varied independently of the kernel structure. This is relevant not only for the application presented here, but also for future work involving geometry refinement, self‑calibration, or more realistic scanner models.
\\\\
Kernel‑level access also provides a natural entry point for future hybrid reconstruction approaches. Physics‑driven operators may be combined with learned components acting as correction terms, regularizers, or adaptive update mechanisms. In this sense, CTrex is not primarily a platform for optimizing a fixed algorithm, but a framework for exploring how physical modelling, numerical discretization, and iterative reconstruction interact.
\\\\
This operator-centric flexibility does, however, come with practical trade‑offs. Kernel‑level programmability increases implementation complexity and requires careful validation to ensure correctness and numerical stability. Additional physical modelling inside the operators may also increase computational cost and memory usage, particularly in schemes involving sub‑voxel sampling or multi‑parameter representations. The framework is therefore not intended as a universal replacement for highly optimized reconstruction software in well‑controlled scenarios, but as an environment for developing and evaluating reconstruction methods that would otherwise be difficult to express.
\\\\
Taken together, these observations suggest that operator‑level accessibility is best understood as an enabling capability for reconstruction research. As acquisition settings become more varied and modelling requirements more application‑specific, the ability to adapt the numerical form of the forward and inverse operators becomes increasingly important for translating new imaging ideas into practically usable reconstruction methods.

\section{Conclusion}\label{sec:conclusion}

This work has presented CTrex as a research‑oriented reconstruction framework that explicitly exposes projection, correction, and backprojection operators as editable computational components within a unified iterative control structure. By bridging high‑level iterative reconstruction strategies with GPU‑level operator implementations, the framework substantially expands the range of imaging scenarios that can be addressed within a single reconstruction environment and enables systematic exploration of operator‑level modelling choices.
\\\\
By treating these operators not as fixed black‑box components but as research objects in their own right, the CTrex framework supports the development of reconstruction strategies for challenging acquisition schemes. The presented application studies demonstrate that editable operators make it possible to investigate how modelling assumptions regarding motion, geometry, sampling density, and spatial prioritization affect reconstruction quality within a consistent reconstruction architecture. Illustrated through several case studies ranging from deformation‑aware reconstruction implemented via kernel‑integrated warping to event‑based 4D reconstruction, as well as geometry‑related applications such as aligned cylindrical coordinates and extended field‑of‑view acquisition.
\\\\
In summary, CTrex offers a practical foundation for developing and evaluating next‑generation micro‑CT reconstruction methods that combine advanced physical modelling, adaptive computation and data‑driven components. By lowering the barrier to modifying core numerical operators, the framework supports methodological innovation and provides a versatile testbed for future research in dynamic and non‑standard computed tomography.


\printcredits

\bibliographystyle{model1-num-names}

\bibliography{cas-refs}

@ARTICLE{dyrect,
  author  = {Goethals, Wannes and Bultreys, Tom and Berg, Steffen and Boone, Matthieu and Aelterman, Jan},
  title   = {DYRECT computed tomography : DYnamic Reconstruction of Events on a Continuous Timescale},
  journal = {IEEE Transactions on Computational Imaging},
  volume  = {11},
  pages   = {638--649},
  year    = {2025},
  doi     = {10.1109/TCI.2025.3566241},
}

@INPROCEEDINGS{WBP,
  author    = {Heyndrickx, Marjolein and Boone, Matthieu and Goethals, Wannes and Bultreys, Tom and Van Hoorebeke, Luc},
  title     = {Weighted backprojection: a reconstruction technique for dynamic micro-CT},
  booktitle = {ICTMS 2019, 4th International Conference on Tomography of Materials and Structures, Abstracts},
  year      = {2019},
  pages   = {176},
  url       = {http://ictms.p.ann.currinda.com/days/2019-07-25/abstract/176},
  location  = {Cairns, QLD, Australia},
}

@ARTICLE{DVC,
  author  = {De Schryver, Thomas and Dierick, Manuel and Heyndrickx, Marjolein and Van Stappen, Jeroen and Boone, Marijn and Van Hoorebeke, Luc and Boone, Matthieu},
  title   = {Motion compensated micro-CT reconstruction for in-situ analysis of dynamic processes},
  journal = {Scientific Reports},
  volume  = {8},
  pages   = {7655},
  year    = {2018},
  doi     = {10.1038/s41598-018-25916-5},
}

@ARTICLE{cylindric,
  author  = {Goethals, Wannes and Heyndrickx, Marjolein and Boone, Matthieu},
  title   = {Fast tomographic inspection of cylindrical objects},
  journal = {Journal of Nondestructive Evaluation},
  volume  = {39},
  number  = {4},
  pages   = {76},
  year    = {2020},
  doi     = {10.1007/s10921-020-00711-3},
}

@PHDTHESIS{thesis_wannes,
  author     = {Goethals, Wannes},
  title      = {{CT} reconstruction in application-driven coordinate systems},
  school     = {Ghent University},
  year       = {2022},
  url        = {http://hdl.handle.net/1854/LU-01GQHVMJQCCCYFK0SW9S9MGPR9},
  eprinttype = {hdl},
  eprint     = {1854/LU-01GQHVMJQCCCYFK0SW9S9MGPR9},
  address    = {Ghent, Belgium},
  note       = {Section 2.5.4: Reconstructing extended field of view CT scans},
}

@INPROCEEDINGS{goethalsStrategiesConeBeam2019,
  author    = {Goethals, Wannes and Heyndrickx, Marjolein and Boone, Matthieu},
  title     = {Strategies in cone beam {CT} inspection of cylindrical objects},
  booktitle = {9th Conference on Industrial Computed Tomography ({iCT} 2019)},
  location  = {Padova, Italy},
  pages     = {8},
  year      = {2019},
  url       = {https://www.ndt.net/article/ctc2019/papers/iCT2019_Full_paper_32.pdf},
}

@ARTICLE{VERLEDEN2021167,
  author  = {Stijn E Verleden and Miranda Kirby and Stephanie Everaerts and Arno Vanstapel and John E McDonough and Erik K Verbeken and Peter Braubach and Matthieu N Boone and Danesh Aslam and Johny Verschakelen and Laurens J Ceulemans and Arne P Neyrinck and Dirk E {Van Raemdonck} and Robin Vos and Marc Decramer and Tillie L Hackett and James C Hogg and Wim Janssens and Geert M Verleden and Bart M Vanaudenaerde},
  title   = {Small airway loss in the physiologically ageing lung: a cross-sectional study in unused donor lungs},
  journal = {The Lancet Respiratory Medicine},
  volume  = {9},
  number  = {2},
  pages   = {167--174},
  year    = {2021},
  doi     = {10.1016/S2213-2600(20)30324-6},
  url     = {https://www.sciencedirect.com/science/article/pii/S2213260020303246},
}

@ARTICLE{cantModelingBlurringEffects2015,
  author  = {Cant, Jeroen and Palenstijn, Willem Jan and Behiels, Gert and Sijbers, Jan},
  title   = {Modeling Blurring Effects Due to Continuous Gantry Rotation: {{Application}} to Region of Interest Tomography},
  journal = {Medical Physics},
  year    = {2015},
  volume  = {42},
  number  = {5},
  pages   = {2709--2717},
  doi     = {10.1118/1.4914422},
  url     = {https://onlinelibrary.wiley.com/doi/abs/10.1118/1.4914422},
}

@ARTICLE{graasASTRAKernelkitGPUaccelerated2024,
  author  = {Graas, Adriaan and Jan Palenstijn, Willem and Van Werkhoven, Ben and Lucka, Felix},
  title   = {{{ASTRA}} Kernelkit: {{GPU-accelerated}} Projectors for Computed Tomography Using Cupy},
  journal = {Applied Mathematics for Modern Challenges},
  year    = {2024},
  volume  = {2},
  number  = {1},
  pages   = {70--92},
  doi     = {10.3934/ammc.2024004},
  url     = {https://www.aimsciences.org//article/doi/10.3934/ammc.2024004},
}

@INPROCEEDINGS{deschryver4DVizualisationInsitu2017,
  author    = {De Schryver, Thomas and Van Stappen, Jeroen and Boone, Marijn A. and Boone, Matthieu and Dierick, Manuel and Cnudde, Veerle and Van Hoorebeke, Luc},
  title     = {{4D} vizualisation of in-situ aluminum foam compression with lab-based motion compensated {X-ray} micro-{CT}},
  booktitle = {7th Conference on Industrial Computed Tomography ({iCT} 2017)},
  location  = {Leuven, Belgium},
  pages     = {6},
  year      = {2017},
  url       = {http://www.ndt.net/events/iCT2017/app/content/Paper/28_DeSchryver.pdf},
}

@ARTICLE{deschryverInlineNondestructiveEvaluation2017,
  author  = {Thomas {De Schryver} and Jelle Dhaene and Manuel Dierick and Matthieu N. Boone and Eline Janssens and Jan Sijbers and Mattias {van Dael} and Pieter Verboven and Bart Nicolai and Luc {Van Hoorebeke}},
  title   = {In-line NDT with X-Ray CT combining sample rotation and translation},
  journal = {NDT \& E International},
  volume  = {84},
  pages   = {89--98},
  year    = {2016},
  doi     = {10.1016/j.ndteint.2016.09.001},
  url     = {https://www.sciencedirect.com/science/article/pii/S0963869516300779},
}

@INPROCEEDINGS{goethalsDVCDynamicCT2019,
  author    = {Goethals, Wannes and De Schryver, Thomas and Van Hoorebeke, Luc and Aelterman, Jan and Boone, Matthieu},
  title     = {{DVC} in dynamic {CT} reconstruction for material characterization},
  booktitle = {{ICTMS} 2019, 4th International Conference on Tomography of Materials and Structures, Abstracts},
  location  = {Cairns, QLD, Australia},
  pages     = {2},
  year      = {2019},
  url       = {http://ictms.p.ann.currinda.com/days/2019-07-23/abstract/169},
}

@ARTICLE{goethalsInsituXrayImaging2020,
  author     = {Goethals, Wannes and Vanbillemont, Brecht and Lammens, Joris and De Beer, Thomas and Vervaet, Chris and Boone, Matthieu},
  title      = {In-Situ {{X-ray}} Imaging of Sublimating Spin-Frozen Solutions},
  journal    = {Materials},
  year = {2020},
  volume     = {13},
  number     = {13},
  doi        = {10.3390/ma13132953},
  url        = {http://hdl.handle.net/1854/LU-8668950},
  eprinttype = {hdl},
  eprint     = {1854/LU-8668950},
}

@ARTICLE{heyndrickxImprovingImageQuality2020a,
  author     = {Heyndrickx, Marjolein and Bultreys, Tom and Goethals, Wannes and Van Hoorebeke, Luc and Boone, Matthieu},
  title      = {Improving Image Quality in Fast, Time-Resolved Micro-{{CT}} by Weighted Back Projection},
  journal    = {Scientific Reports},
  year = {2020},
  volume     = {10},
  number     = {1},
  doi        = {10.1038/s41598-020-74827-x},
  url        = {http://hdl.handle.net/1854/LU-8679712},
  eprinttype = {hdl},
  eprint     = {1854/LU-8679712},
}

@INPROCEEDINGS{heyndrickxImprovingReconstructionDynamic2016,
  author    = {Heyndrickx, Marjolein and De Schryver, Thomas and Dierick, Manuel and Boone, Matthieu and Bultreys, Tom and Cnudde, Veerle and Van Hoorebeke, Luc},
  title     = {Improving the reconstruction of dynamic processes by including prior knowledge},
  booktitle = {{HD-Tomo-Days}, Abstracts},
  location  = {Lyngby, Denmark},
  year      = {2016},
}

@ARTICLE{manigrassoMultiframeDVCTemporal2022,
  author     = {Manigrasso, Zaira and Goethals, Wannes and Moazami Goudarzi, Niloofar and Boone, Matthieu and Samaro, Aseel and Vervaet, Chris and Philips, Wilfried and Aelterman, Jan},
  title      = {Multi-Frame {{DVC}} for Temporal Image Sequences},
  journal    = {Frontiers in Materials},
  year = {2022},
  volume     = {9},
  doi        = {10.3389/fmats.2022.998311},
  url        = {http://hdl.handle.net/1854/LU-8769963},
  eprinttype = {hdl},
  eprint     = {1854/LU-8769963},
}

@ARTICLE{quintanaDigitalVolumeCorrelation,
  author  = {Quintana, Enrico C. and Reu, Phillip L. and Jimenez, Edward and Thompson, Kyle R. and Kramer, Sharlotte},
  title   = {Volumetric Digital Image Correlation for Materials Characterization},
  journal = {e-Journal of Nondestructive Testing},
  volume  = {21},
  number  = {7},
  year    = {2016},
  url     = {https://www.ndt.net/?id=19219},
  note    = {Presented at the 19th World Conference on Non-Destructive Testing, Munich, Germany, 13--17 June 2016},
}

@ARTICLE{aarleFastFlexibleXray2016a,
  author  = {Wim van Aarle and Willem Jan Palenstijn and Jeroen Cant and Eline Janssens and Folkert Bleichrodt and Andrei Dabravolski and Jan De Beenhouwer and K. Joost Batenburg and Jan Sijbers},
  title   = {Fast and flexible X-ray tomography using the ASTRA toolbox},
  journal = {Opt. Express},
  volume  = {24},
  number  = {22},
  pages   = {25129--25147},
  year    = {2016},
  month   = {Oct},
  doi     = {10.1364/OE.24.025129},
  url     = {https://opg.optica.org/oe/abstract.cfm?URI=oe-24-22-25129},
}

@ARTICLE{biguriTIGREV3Efficient2025,
  author  = {Biguri, Ander and Sadakane, Tomoyuki and Lindroos, Reuben and Liu, Yi and Sabaté Landman, Malena and Du, Yi and Lohvithee, Manasavee and Kaser, Stefanie and Hatamikia, Sepideh and Bryll, Robert and Valat, Emilien and Wonglee, Sarinrat and Blumensath, Thomas and Schönlieb, Carola-Bibiane},
  title   = {{{TIGRE}} v3: {{Efficient}} and Easy to Use Iterative Computed Tomographic Reconstruction Toolbox for Real Datasets},
  journal = {Engineering Research Express},
  year = {2025},
  volume  = {7},
  number  = {1},
  pages   = {015011},
  doi     = {10.1088/2631-8695/adbb3a},
}

@ARTICLE{jorgensenCoreImagingLibrary2021,
  author  = {Jørgensen, J. S. and Ametova, E. and Burca, G. and Fardell, G. and Papoutsellis, E. and Pasca, E. and Thielemans, K. and Turner, M. and Warr, R. and Lionheart, W. R. B. and Withers, P. J.},
  title   = {Core {{Imaging Library}} - {{Part I}}: A Versatile {{Python}} Framework for Tomographic Imaging},
  journal = {Philosophical Transactions of the Royal Society A: Mathematical, Physical and Engineering Sciences},
  volume  = {379},
  year = {2021},
  number  = {2204},
  pages   = {20200192},
  doi     = {10.1098/rsta.2020.0192},
  url     = {https://royalsocietypublishing.org/doi/10.1098/rsta.2020.0192},
}

@ARTICLE{papoutsellisCoreImagingLibrary2021,
  author  = {Papoutsellis, Evangelos and Ametova, Evelina and Delplancke, Claire and Fardell, Gemma and Jørgensen, Jakob S. and Pasca, Edoardo and Turner, Martin and Warr, Ryan and Lionheart, William R. B. and Withers, Philip J.},
  title   = {Core {{Imaging Library}} - {{Part II}}: Multichannel Reconstruction for Dynamic and Spectral Tomography},
  journal = {Philosophical Transactions of the Royal Society A: Mathematical, Physical and Engineering Sciences},
  volume  = {379},
   year = {2021},
  number  = {2204},
  pages   = {20200193},
  doi     = {10.1098/rsta.2020.0193},
  url     = {https://royalsocietypublishing.org/doi/10.1098/rsta.2020.0193},
}

@ARTICLE{multi_res,
  author       = {De Witte, Yoni and Vlassenbroeck, Jelle and Van Hoorebeke, Luc},
  title        = {A multiresolution approach to iterative reconstruction algorithms in {X}-ray computed tomography},
  journal      = {IEEE Transactions on Image Processing},
  volume       = {19},
  number       = {9},
  pages        = {2419--2427},
  year         = {2010},
  doi          = {10.1109/TIP.2010.2046960},
  url          = {http://doi.org/10.1109/TIP.2010.2046960},
  issn         = {1057-7149},
  language     = {eng},

}

@ARTICLE{ritReconstructionToolkitRTK2014,
  author  = {Rit, S and Vila Oliva, M and Brousmiche, S and Labarbe, R and Sarrut, D and Sharp, G C},
  title   = {The {{Reconstruction Toolkit}} ({{RTK}}), an Open-Source Cone-Beam {{CT}} Reconstruction Toolkit Based on the {{Insight Toolkit}} ({{ITK}})},
  journal = {Journal of Physics: Conference Series},
  year = {2014},
  volume  = {489},
  number  = {1},
  pages   = {012079},
  doi     = {10.1088/1742-6596/489/1/012079},
}

@ARTICLE{tuyInversionFormulaConeBeam1983,
  author  = {Tuy, Heang K.},
  title   = {An {{Inversion Formula}} for {{Cone-Beam Reconstruction}}},
  journal = {SIAM Journal on Applied Mathematics},
  year = {1983},
  volume  = {43},
  number  = {3},
  pages   = {546--552},
  doi     = {10.1137/0143035},
  url     = {https://epubs.siam.org/doi/10.1137/0143035},
}

@ARTICLE{josephFastRayTracing1982,
  author  = {Joseph, Peter M.},
  title   = {An Improved Algorithm for Reprojecting Rays through Pixel Images},
  journal = {IEEE Transactions on Medical Imaging},
  volume  = {1},
  number  = {3},
  pages   = {192--196},
  year    = {1982},
  doi     = {10.1109/TMI.1982.4307572},
}

@ARTICLE{smithImageReconstructionConeBeam1985,
  author  = {Smith, Bruce D.},
  title   = {Image {{Reconstruction}} from {{Cone-Beam Projections}}: {{Necessary}} and {{Sufficient Conditions}} and {{Reconstruction Methods}}},
  journal = {IEEE Transactions on Medical Imaging},
  year = {1985},
  volume  = {4},
  number  = {1},
  pages   = {14--25},
  doi     = {10.1109/TMI.1985.4307689},
  url     = {https://ieeexplore.ieee.org/abstract/document/4307689},
}



\end{document}